\documentclass[preprint]{aastex631}
	
\usepackage{soul}
\usepackage{hyperref}
\usepackage{url}
\usepackage{times}
\usepackage{graphicx}
\usepackage{amsfonts}
\usepackage{import}
\usepackage{listings}
\usepackage{appendix}
\usepackage{mathtools}
\usepackage{amsmath}
\usepackage{aas_macros}

\usepackage{calligra}
\usepackage{natbib}

\begin{document}

\bibliographystyle{aasjournal}     

\title{The dynamics of self-gravity wakes in the Mimas 5:3 bending wave: modifying the linear theory}
\author{Daniel D. Sega}
\email{daniel.sega@colorado.edu}
\affiliation{Laboratory for Atmospheric and Space Physics, University of Colorado at Boulder, Boulder, CO, 80309, USA}
\author{Glen Stewart}
\affiliation{Laboratory for Atmospheric and Space Physics, University of Colorado at Boulder, Boulder, CO, 80309, USA}
\author{Joshua E. Colwell}
\affiliation{Department of Physics, University of Central Florida, Orlando, FL, 32816, USA}
\author{Girish M. Duvvuri}
\affiliation{Laboratory for Atmospheric and Space Physics, University of Colorado at Boulder, Boulder, CO, 80309, USA}
\affiliation{Center for Astrophysics and Space Astronomy, University of Colorado at Boulder, Boulder, CO, 80309, USA}
\author{Richard Jerousek}
\affiliation{Department of Physics, University of Central Florida, Orlando, FL, 32816, USA}
\author{Larry Esposito}
\affiliation{Laboratory for Atmospheric and Space Physics, University of Colorado at Boulder, Boulder, CO, 80309, USA}
\date{\today}
\begin{abstract}
  \noindent  The satellite Mimas launches a bending wave ---a warping of the rings that propagates radially through self-gravity--- at the 5:3 inner vertical resonance with Saturn's rings. We present a modification of the linear bending wave theory \citep{SCL} which includes the effects of satellite self-gravity wakes on the particles in the wave. We show that, when treated as rigid, these wakes generate an extra layer of particles whose number density is proportional to the magnitude of the slope of the warped ring. Using a ray-tracing code we compare our predictions with those of \citet{SCL} and with 60 stellar occultations observed by the Cassini Ultraviolet Imaging Spectrograph (UVIS) and find that the extra layer of particles of our perturbed bending wave model has a considerable explanatory power for the UVIS dataset. Our best model explains the most discrepant and surprising features of the Mimas 5:3 bending wave; the enhancement of the signal for the cases of occultations with high ring opening angle and the bigger-than-expected viscosity, $\nu = 576 \, \mathrm{cm^2/s}$, which is more than double the viscosity computed from density waves \citep{Tiscareno}. This shows that self-gravity wakes can be effective at transporting angular momentum in a vertically perturbed disk. Relative to neighboring density waves \citep{Tiscareno}, we find a lower-than-expected value for the surface mass density, $\sigma = 36.7 \, \mathrm{g/cm^2}$, which suggests that the enhanced viscous interactions may be transporting material into the surrounding regions.

\end{abstract} 

\section{Introduction} \label{sec1}

\begin{figure}[t]
  \centering
  \includegraphics[width = 0.9\linewidth]{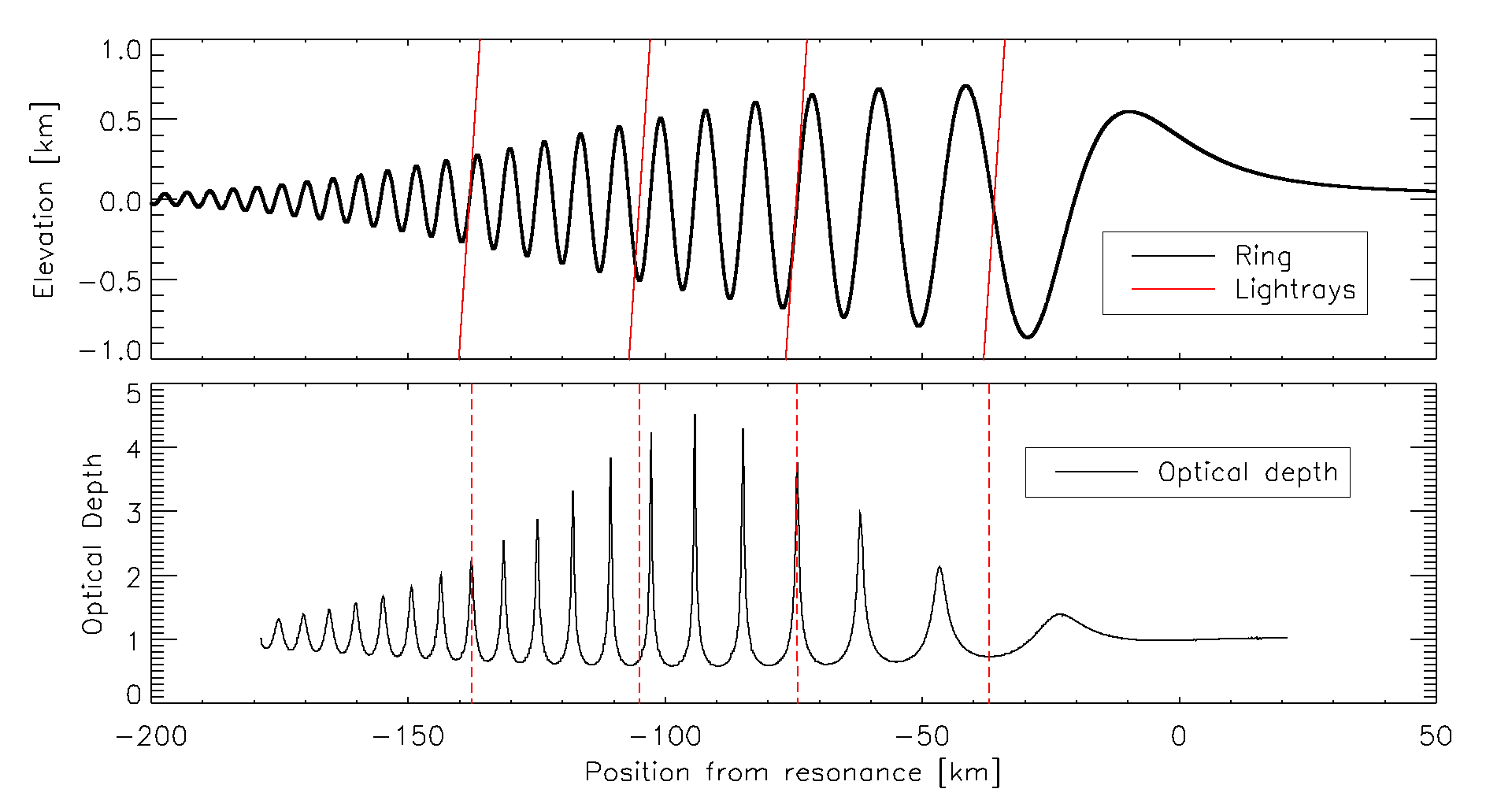}
  \caption{Theoretical predictions for the vertical displacement (top panel) and apparent optical depth (bottom panel) according to SCL. The red rays in the top panel represent light rays that are being transmitted through the ring, and the corresponding optical depth probed for each light ray is marked by the red dashed lines in the bottom panel. The angle between the mean ring plane and the incident light rays is $B = 26^{\circ}$ (note that the axes do not have the same scale), and the angle between the radial direction and the light rays projected onto the plane is $\phi =0^{\circ}$. We can see how the light ray goes through more material when it aligns with the ring's slope (which corresponds to a peak in optical depth), and through less material when it travels perpendicular to the slope (which corresponds to a trough in optical depth).}
  \label{fig:1}
\end{figure}

A bending wave (BW) is a warping of a disk (a circular and thin mass plane) caused by perturbations normal to the disk that propagates due to self-gravity. Warped disks are ubiquitous in astrophysics and can be found around young stars \citep{becker}, black holes \citep{mnras2021}, and around planets \citep{SCL}, for which we have a vast amount of observations available. While warps are only understood to a limited extent (for instance, see \citealt{Nelson}), they are known to occur when out-of-the-plane forces perturb a disk that orbits a central body. In the case of the rings of Saturn, the out-of-the-plane force is produced when satellites with inclined orbits perturb the ring particles at a frequency that coincides with a rational multiple of these particles' vertical motion at some radial distance from Saturn \citep{Murray2000}. Many of these resonances occur in the A ring of Saturn, where we focus on the 5:3 vertical resonance due to the moon Mimas at $131902 \,\mathrm{km}$ from Saturn's center.  A radial cut through an idealized bending wave is shown on the top panel of Figure \ref{fig:1}.


\citet*{SCL} (henceforth SCL) were the first to confirm the presence of bending waves in the rings by adapting the theory of warped galaxies developed by \citet{1980AA} and analyzing images from Voyager I. While there was general agreement with the SCL theory for weak bending waves, the Mimas 5:3 BW proved difficult to fit. Later, \citet{Lissauer} focused on this specific wave and obtained a value for the viscosity by using similar methods. A general study of waves using occultations from Voyager's photopolarimeter was then done by \citet{ESPOSITO1}, while \citet{gresh} did an extensive study focusing on bending waves using low opening angle radio occultations. Neither managed to explain the shape of the Mimas 5:3 BW, and \citet{gresh} suggested that the damping mechanism for this wave needs revision. On the theoretical side, \citet{Chakrabarti1988, Chakrabarti1989} considered radial shear of particles as a damping mechanism for the Mimas 5:3 BW but it yielded an inconsistent result for the viscosity. All of these efforts were based on the Voyager observations; however, Cassini data has yet to be used to improve the theory of bending waves.

\begin{figure}[t]
  \centering
    \includegraphics[width = 0.9\linewidth]{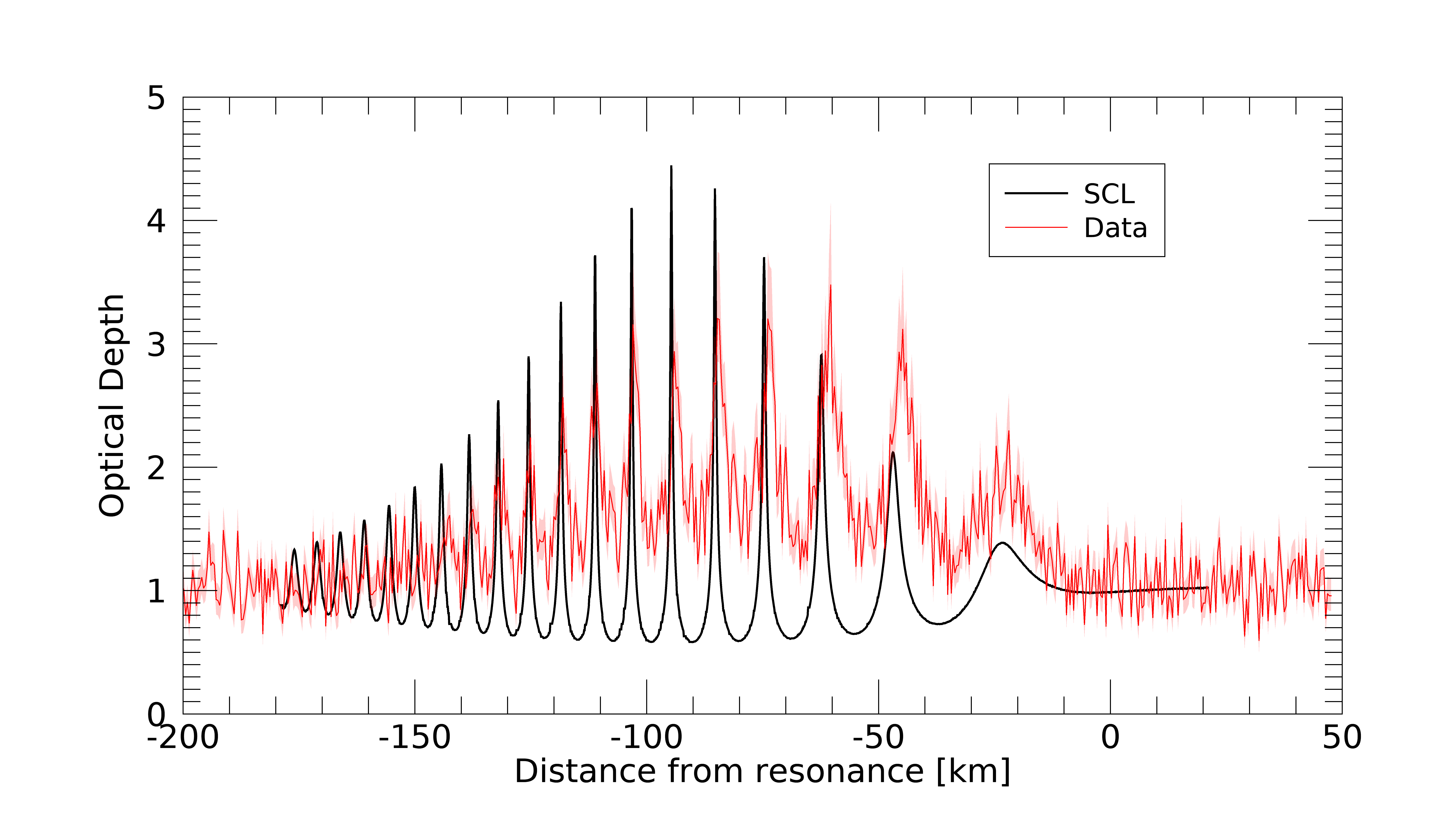}
  \caption{We compare an occultation of $\gamma$-Pegasus (rev 32I, $B = 20.3^{\circ}$, $B_{\mathrm{eff}} = 26^{\circ}$) seen by Cassini (red) to the standard bending wave theory (black). The resonance occurs at a radius of $131902 \, \mathrm{km}$ and the wave propagates towards Saturn (left of the plot). We see three major discrepancies: (1) The SCL theory overpredicts the peaks of the optical depth and underpredicts the troughs. (2) The wavelengths do not match for the first three cycles. (3) The theory underpredicts how fast the wave damps, as peaks are still predicted to appear around $-150 \, \mathrm{km}$, but the observed wave gets buried in the noise before that point. Using the viscosity of $\nu = 260 \, \mathrm{cm^2/s}$ from \citet{Lissauer}'s analysis of this wave and \citet{Tiscareno}'s analysis of the Prometheus 11:12 density wave.}
  \label{fig:2}
\end{figure}

\begin{figure}[t]
  \centering
  \includegraphics[width= 0.9\linewidth]{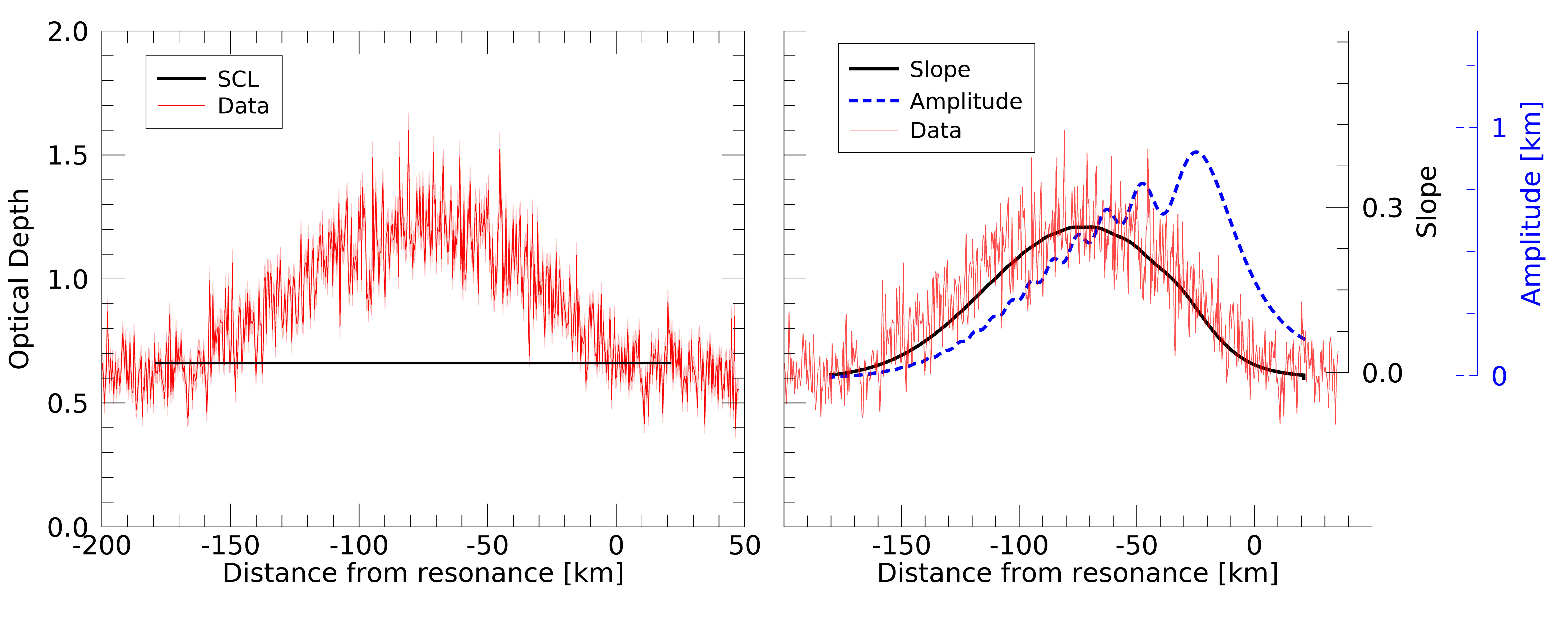}
  \caption{Left panel: we compare an occultation of $\beta$-Centauri (rev85I, $B = 66.7^{\circ}$, $B_{\mathrm{eff}} = 87^{\circ}$) (solid red), to the SCL model (solid back). The error bars are displayed as a light red filled curve in the background of the plot. The SCL theory predicts a uniform optical depth; however, the data shows a symmetric rise in the optical depth centered at $-80 \, \mathrm{km}$ from resonance. Right panel: the same optical depth profile of $\beta$-Centauri (solid red) is now compared to the theoretical (SCL theory) maximum slope of the wave (solid black) and the amplitude of the wave (dashed blue) for a viscosity of $566 \,  \mathrm{cm^2/s}$ and a ring surface density of $\sigma = 36.3 \, \mathrm{g/cm^2}$; these values are the best-fit wave parameters of the model presented in this paper. }
  \label{fig:3}
\end{figure}

Figure \ref{fig:1} shows how a BW can influence the optical depth of the ring measured by stellar occultations (which are photometric measurements of a star as it is occulted by the rings) assuming the ring behaves as predicted by the SCL theory. The red dashed lines in the bottom panel show the optical depths corresponding to the light rays in the top panel: the optical depth maxima (minima) occur when the light rays are the most parallel (perpendicular) to the local warped ring. From looking at Cassini occultations (Figures \ref{fig:2} and \ref{fig:3}), and by comparing to \citet{gresh}, we have identified 3 problems that appear when comparing SCL with the data:

  \textit{(1) Wave profile.} \\
  \indent The poor prediction of the wave’s morphology can be observed in Figure \ref{fig:2}, where we show a plot of the optical depth of an occultation of the star $\gamma$-Pegasus measured by UVIS (red line) compared to SCL theory (black line). The effective opening angle of an occultation is defined as $B_{\mathrm{eff}}=\tan^{-1}(\tan B/\cos \phi)$ where $B$ is the angle between the mean ring plane and the incident light rays, and $\phi$ the angle between the radial direction and the light rays, projected onto the plane (see \citealt{gresh}); the lower $B_{\mathrm{eff}}$ is, the higher the variations in the optical depth. As shown in Figure \ref{fig:1}, the peaks and troughs of this optical depth depend on the inclination, or the slope, of the ring relative to the line-of-sight to the star. The overprediction of the optical depth peak interior to $-80 \, \mathrm{km}$ seen in Figure \ref{fig:2} suggests that the SCL theory overestimates this inclination.

  SCL also predicts that all variation of the optical depth of the disk relative to the flat ring is caused only by the inclination of the ring; therefore, in occultations where the line-of-sight is normal to the rings, no variation of optical depth is expected. In the left panel of Figure \ref{fig:3}, however, the data shows otherwise. An approximately normal (to the rings) occultation still shows a symmetric variation of the optical depth, centered about $80 \, \mathrm{km}$ inside the resonance location, which suggests an extra layer of material being generated within the BW that may be caused by the fragmentation of particles in the ring \citep{gresh}. The right panel of Figure \ref{fig:3} shows that the position of the peak and the overall shape of this optical depth enhancement is correlated with the theoretical prediction for the maximum \textit{slope} of the wave (solid black line) instead of the maximum elevation (dashed blue line), which further suggests that this fragmentation is higher where the slope of the ring is steeper. In this work, we suggest a plausible physical mechanism that explains this extra layer of particles and its apparent correlation with the slope of the wave.

  This suggested extra layer of particles could be described as a haze, as a comparison between the Visible and Infrared Mapping Spectrometer (VIMS) and UVIS data (see Table \ref{uvisvims}) shows the presence of a small amount of sub-mm size particles which diffract the infrared light out of the detector increasing the infrared optical depth \citep{Jerousek}. This is an uncommon sight in the A-ring as sub-mm particles normally are rapidly accreted into bigger particles \citep{bodrova, HARBISON}, further motivating a modification of SCL theory that accounts for the disruption of material in the BW region.

  \begin{table}[h]
  \centering
  \begin{tabular}{ |c|| c| c| }
    \hline

           & $\lambda \; [\mathrm{nm}]$  & $\frac{\bar{\tau}_{\mathrm{wave}}}{\bar{\tau}_{\mathrm{bg}}} \; \mathrm{[-]}$    \\
  \hline
  \hline
  
  VIMS      &  $2886 - 2977$                                     &  $1.520 + 0.026 \, / - 0.025  $ \\

  UVIS      &  $110 - 190$                                       &   $1.352 + 0.028 \,/ - 0.027 $ \\
  \hline

\end{tabular}
\caption{Simultaneous Infrared and Ultraviolet occultations of $\alpha$-Scorpius taken by the VIMS and UVIS instruments. VIMS optical depth is larger by $13\%$ compared to UVIS. For $\tau_{\mathrm{bg}}$ we averaged data between $131600$ and $131700 \, \mathrm{km}$ and for $\tau_{\mathrm{wave}}$ we averaged data between $131862$ and $131762 \, \mathrm{km}$ }
\label{uvisvims}
\end{table}

\textit{(2) Wavelength near resonance.} \\
\indent Figure \ref{fig:2} indicates that, while the theoretical wavelength works fairly well at later wave crests (this wave propagates to the left in the plot), it differs from observed values near resonance, a result seen in \citet{gresh}. The extra layer of particles suggested by Figure \ref{fig:3} and Table \ref{uvisvims} may produce a radial change in surface density---which affects the local wavelength (SCL; \citealt{lehmann})---or modify the shape of the first optical depth peaks. These effects may aid in explaining the observed wavelength mismatch.

\textit{(3) Viscosity.} \\
   \indent  The ring’s viscosity controls the shape and amplitude of the peaks of the optical depth profile seen in Figure \ref{fig:2}. Viscous interactions reduce the vertical displacement of ring particles; which, in turn, changes the maximum extinction at the point where the ring’s slope and the line-of-sight coincide. In other words, the peak amplitude of the optical depth pattern depends on the viscosity. In Figure \ref{fig:2}, the SCL model (black line) predicts a highly peaked optical depth profile and a wave that propagates beyond the observed variations seen in the data (red line). This suggests that the fluid shear description of the damping is inadequate, and/or the values suggested by previous Mimas 5:3 BW studies \citep{Lissauer, gresh}, and Cassini density waves studies \citep{Tiscareno}, are too low to describe this wave. Regardless, the observed peaks of the optical depth and how fast the wave damps indicate that the current theory underpredicts the amount of energy and momentum diffused in the wave region. This missing diffusion may be caused by the same mechanism that produces the extra layer of particles suggested by Figure \ref{fig:3}.

    All three issues can be explored in depth using Cassini ring occultation data. Cassini observed the Mimas 5:3 BW in $217$ occultations with the UVIS High Speed Photometer (HSP) \citep{Esposito2004} out of which $130$ have a signal-to-noise high enough to detect the variations in the data due to the bending wave. Moreover, using Cassini data, \citet{Colwell2006} confirmed the existence of \textit{self-gravity wakes}, which are elongated clumps of particles that periodically disaggregate, and are $10$ to $100$ times bigger than a typical A ring particle ($\sim 1 \, \mathrm{m}$) \citep{Jerousek, salo2018}. The consequences of adding self-gravity wakes to the SCL theory are derived in \S \ref{sec:2} via the assumption that they react to the relevant torques as rigid bodies; here we present a mechanism that explains the extra optical depth signal seen in normal occultations. In \S \ref{sec:3} we describe the transmission model, our selection criteria for the $60$ UVIS occultations used, and how the data were reduced. In \S \ref{sec:4} we describe the ray-tracing code. In \S \ref{sec:5} we compare our bending wave model to the data followed by a discussion in \S \ref{sec:6}, where we consider how our model addresses the above problems. In \S \ref{sec:7} we summarize our conclusions.

{\section{Self-gravity wakes and the Bending wave}

\label{sec:2}

Self-gravity wakes are clumps of particles with an elongated structure at an angle $\theta_{w}$ from the azimuthal direction (see Figure \ref{wakes}), which are held together by self-gravity against the Keplerian shear, and whose constituent particles present a resistance to changing their relative distances, $\Delta r_{ij}\approx 0$ \citep{Daisaka1999, Yuxi}. In other words, particles within the wakes show coherent motion, causing the wake to keep a consistent shape and orientation in timescales of one orbital period \citep{Salo2004}. Self-gravity wakes were first hypothesized in the galactic context \citep{1964AJ} as non-axisymmetric disturbances generated by the interplay of the self-gravity of a rotating disk and its shear. They were first proposed to exist in the rings by \citet{colombo}, and \citet{dandp} and \citet{Dunn} modeled and corroborated their effects on Voyager images and Very Large Array observations, respectively. They were first numerically modeled for planetary rings by \citet{salo1992} and were detected in Saturn's rings in \citet{Colwell2006} as an azimuthal asymmetry in the optical depth of UVIS occultations.

The SCL linear bending wave theory takes the rings to be smooth and of uniform density, so the presence of these structures was not accounted for in the dynamics. To study the effects of self-gravity wakes we assume that they remain rigid at least during the time it takes the wave to advance one wavelength in the frame co-rotating with the wake, which is about one orbital period (see Eqn. \ref{h} below). At least some self-gravity wakes are expected to remain together for that long \citep{Salo2004}: simulations even suggesting possible lifetimes of many orbital periods \citep{michikoshi}.

\begin{figure}[t]
  \centering
  \includegraphics[width= 0.7\linewidth]{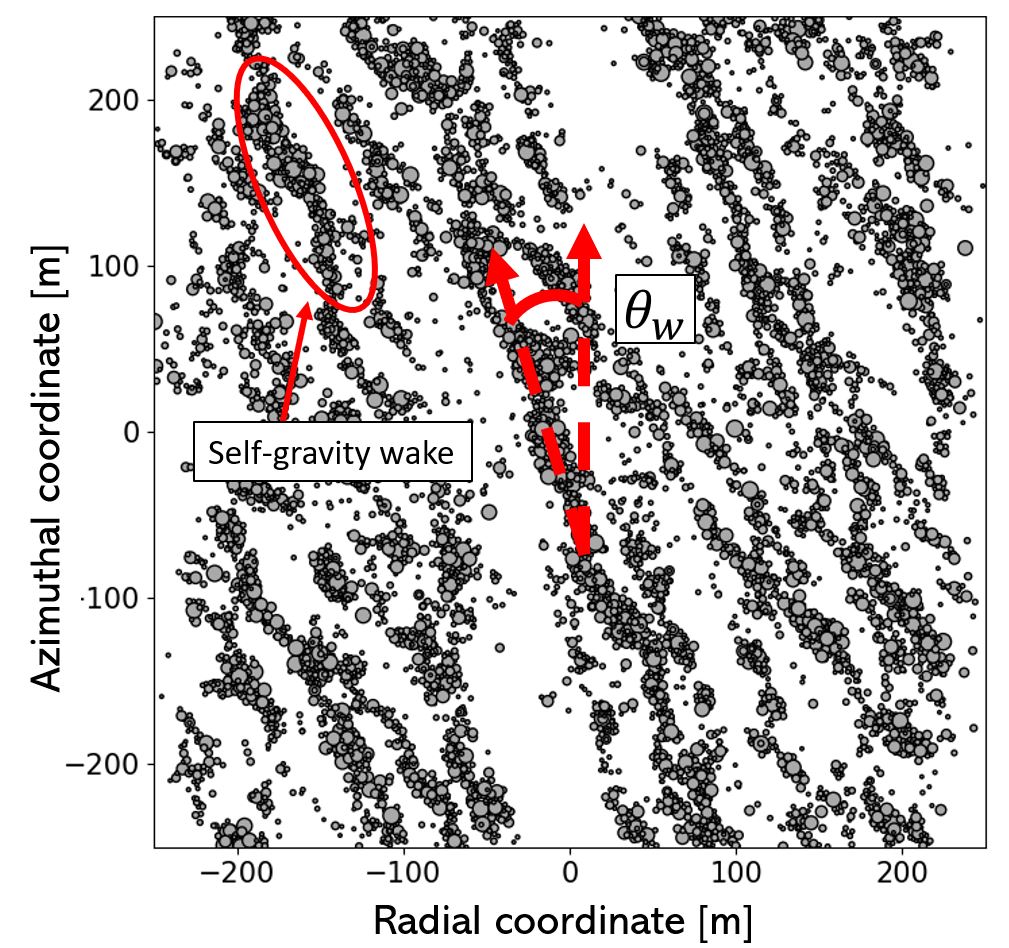}
  \caption{Snapshot of self-gravity wakes in a shear-box simulation produced with the shear-sheet \texttt{REBOUND} example \citep{rebound}, for visualization purposes only. Circled in red we can see an oriented self-gravity wake; we can also see they tend to point at an angle $\theta_{w}$ with the azimuthal.}
  \label{wakes}
\end{figure}

\begin{figure}
  \centering
  \includegraphics[width= 1\linewidth]{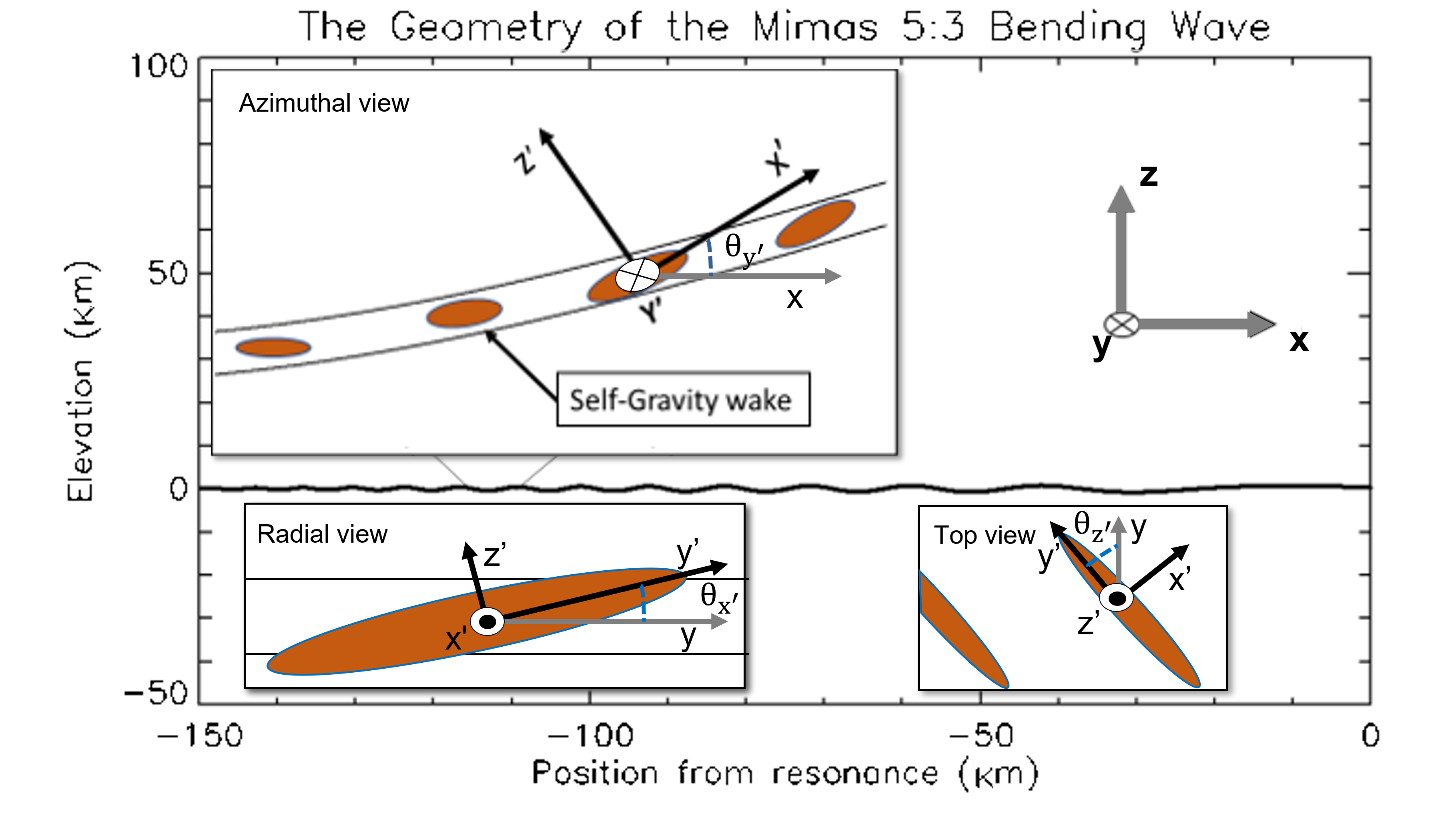}
  \caption{Bending wave (to scale) and the two frames of reference used in this work. The frame on the right is centered at resonance ($x=0$ at resonance), co-rotating and defined by the unprimed axis $(\hat{x}, \hat{y}, \hat{z})$ while the Eulerian frame on the wakes (top left) is centered on the center of mass of a self-gravity wake at a distance $r$ from resonance and it is defined by the primed coordinates $(\hat{x'},\hat{y'},\hat{z'})$. We also represent here the angles $\theta_{q'}=\int\omega_{q'}dt$  where  $\omega_{q'}$ is the angular velocity of a rotation about the $q'$-axis where $q'$ can be $x', y',$ or $z'$ (the principal axes of the wake). The angles are $0$ when the prime and unprimed axis align.}
  \label{frames}
\end{figure}

Within the BW the wakes are subjected to different torques that affect their orientation ($\theta_{q'}$, see Figure \ref{frames}) as they oscillate in elevation ($z$) with the wave; these torques and their effects on the rotational motion of the wakes are computed in \S \ref{sec:21}. We will see that, as the wakes change their orientation, the wake particles at some distance from the center of the wake will gain a difference in vertical speed with respect to the surrounding ring particles (which are assumed to follow the expected SCL vertical motion), resulting in collisions of order $1 \, \mathrm{cm/s}$ in relative velocities which release material residing on the wake's surface. The predicted properties of the resultant haze are described in \S \ref{sec:23}.

\subsection{The torques that couple the wake's rotation and elevation}

Assuming the wakes remain rigid, we can compute their rotation by solving Euler's equations of rigid motion:

\label{sec:21}
\begin{equation}
  \dot{\omega}_{x'} = \frac{\tau_{x'}}{I_{x'}}+\frac{I_{y'}-I_{z'}}{I_{x'}} \omega_{y'} \omega_{z'}
    \label{eulerx}
\end{equation}

\begin{equation}
  \dot{\omega}_{y'} = \frac{\tau_{y'}}{I_{y'}}+\frac{I_{z'}-I_{x'}}{I_{y'}} \omega_{z'} \omega_{x'}
   \label{eulery}
\end{equation}

\begin{equation}
  \dot{\omega}_{z'}= \frac{\tau_{z'}}{I_{z'}} + \frac{I_{x'}-I_{y'}}{I_{z'}} \omega_{x'} \omega_{y'}
  \label{eulerz}
\end{equation}

\noindent where $I_{q'}$, $\omega_{q'}$, and $\tau_{q'}$ are the moment of inertia, angular velocity, and torques, respectively, about the principal axis $q'$ (where $q'=x'$, $y'$, or $z'$).

To solve these equations we need to compute the torques about each principal axis ($\tau_{x'}$, $\tau_{y'}$, $\tau_{z'}$), which will depend, among other things, in the elevation of the center of mass (CM) of the wake, effectively coupling the rotation of the wake with its vertical position. To compute these torques we set the reference frame in the center of mass of a wake a distance $x$ from resonance (see Figure \ref{frames}), and take the time dependence of the vertical coordinate, $z$, of the CM to act harmonically in time with a frequency Doppler-shifted by the orbit of the wake:

\begin{equation}
  \omega'=m(\Omega_M - \Omega) + \mu_M
  \label{omega}
\end{equation}

 \noindent where $\Omega$ is the mean motion frequency of the wake's CM,  $\Omega_M$ and $\mu_M$ are the mean motion and vertical frequencies of Mimas, and $m$ is the azimuthal wave number ($m=4$ for our case). This is in accord with the SCL linear bending wave theory translated to a co-rotating frame represented by the unprimed frame $(x, y, z)$ in Figure \ref{frames}, where $x$ and $y$ are the radial and azimuthal coordinates respectively. We also introduce angle coordinates which represent rotations about the principal axis of rotation $(x',y',z')$ of the wake and for concreteness consider the wake to be a parallelopiped (or a `granola bar') with sides $(W, L, H) = (18 \, \mathrm{m}, 232 \, \mathrm{m}, 4 \, \mathrm{m})$ accordingly. We extract these dimensions from transmission models of the A ring \citep{hedman,Jerousek}, and numerical simulations \citep{salo2018} where  $H/W \sim 0.2$, $L \sim 4\lambda_T$ and $W \sim \lambda_T/3$; $\lambda_T \equiv 4\pi^2G\sigma/\Omega^2 \sim 58  \, \mathrm{m}$ is the Toomre critical wavelength \citep{1964AJ}, where $\sigma$ is the surface mass density and $\Omega$ the mean motion frequency.

The motion of a wake particle at a point $(x',y',z')$ with respect to the unprimed co-rotating frame will then be given by adding the CM velocity (whose motion is governed by the SCL theory) to the velocity due to the wake's rotation, obtained from Eqns. (\ref{eulerx}), (\ref{eulery}), and (\ref{eulerz}). The equation for the z-coordinate for the CM of a wake a position $x$ from resonance as a function of time $t$ is given by:

\begin{equation}
  z(x,t) = \mathrm{Re}\big[h(x)e^{i(\omega' t)}\big],
  \label{h}
\end{equation}

\[h(x) = \mathfrak{H}A_V e^{-(\frac{x}{\chi_D(\nu)})^{3}}\]

\noindent where $\mathfrak{H}$ is a complex Fresnel integral, $A_V$ is the wave amplitude (see Eqn. 27 in SCL), $\chi_D$ is the damping length, which depends on the kinematic viscosity $\nu$ (see Eqn. 98 in \citealt{1984}), and where $\omega'$ is the Doppler-shifted forcing frequency. While for $kx >> 1$, the wave profile $h(x)$ takes a WKBJ form $h(x) \sim A(r)e^{\int ik dx}$, we will use the exact expressions for the amplitude ($\sqrt{hh^*}$) and slope's magnitude ($\sqrt{\frac{dh}{dr} (\frac{dh}{dr})^*}$). For the explicit form of $h(x)$ see \citet{segatesis}, Eqn. (C.54).

Note that we have eliminated $z$'s usual explicit dependence on azimuthal position ($\Theta$) by using the Doppler-shifted frequency $\omega'$, which amounts to making the substitution $\Theta$ = $\Omega t$ in Eqn. (26) of SCL. In other words, we are moving from an Eulerian to a co-rotating Lagrangian description of the rings. Thus, the Lagrangian derivative used in SCL, $D/Dt=\partial/\partial t + \Omega \partial/\partial \Theta$, becomes the time derivative $\partial/\partial t$ in the co-rotating frame.

Knowing the motion of the CM of a wake at a distance $x$ from resonance, we proceed to compute the relevant torques on the wake. We first consider the torques caused by tidal forces and the local environment of the BW, while ignoring the gravitational effects of other wakes. These amount to: the tidal torque due to Saturn ($\tau_{\mathrm{tidal}}$, \S \ref{subsub3}), the torque due to the Keplerian shear ($\tau_{\mathrm{Kep}}$, \S \ref{subsub4}), the acceleration torque due to the BW ($\tau_{\mathrm{BWacc}}$, \S \ref{subsub1}), and the torque due to the vertical shear of the BW ($\tau_{\mathrm{BWsh}}$, \S \ref{subsub2}). These torques cause the most important aspects of the wake's rotation, which we analyze in the idealized case of an isolated wake (\S \ref{subsub5}). Later, we derive the torque due to the wake-to-wake gravitational interaction ($\tau_{\mathrm{wake}}$, \S \ref{subsub6}) and study its effect (\S \ref{subsub7}).

\subsubsection{Tidal torque}
\label{subsub3}

We begin by deriving the torques about the $z'$-axis, which are also present outside the BW region. We assume that the wake's inclination outside the ring's mean plane remains small, so that $\theta_{x'},\theta_{y'} < 15^{\circ}$ and $\theta_{z'}$ represents the pitch-angle of the self-gravity wake (in Appendix \ref{AA} we show the general expression for an arbitrary wake orientation). The torques about the $z'$-axis are the tidal torque---which tends to align the wake's long-axis with the radial direction---and the Keplerian shear torque that aligns the wake with the azimuthal. Later, we will see how the BW torques change this picture.

We start with the tidal force on a piece of the wave with mass $dm$ \citep{Murray2000} \[F_{\mathrm{tidal}}=3\frac{GM_S}{a^3}y'*\sin\theta_{z'}\cos\theta_{z'}dm\] where $a$ is the distance from Saturn's center, $G$ the gravitational constant, $M_s$ the mass of Saturn and $y'$ the coordinate along the long-axis of the wake. Using $dm=HW\rho_{w}dy'$ for the differential mass of the wake (where $\rho_{w}$ is the wake's density) and integrating with respect to $y'$ we arrive at the tidal torque:

  \begin{equation}
    \tau_{\mathrm{tidal};z'} = 2\frac{GM_S}{a^3}\sin{\theta_{z'}}\cos{\theta_{z'}}\left(\frac{L}{2}\right)^3*\rho_{\mathrm{Roche}} HW
    \label{tidal}
  \end{equation}

  For the density of the wake ($\rho_w$) we will use $\rho_{\mathrm{Roche}}$, defined as the bulk-density of an object that fills its Hill sphere \citep{Tiscareno2013}. Aggregates within the rings, such as moonlets, have been shown to have this density within a $20\%$ margin \citep{porco}, and numerical simulations have shown similar results for self-gravity wakes \citep{salo2018}. Notice that this torque is positive when $0 < \theta_{z'} < 90^{\circ}$ and becomes negative when $\theta_{z'} > 90^{\circ}$, causing the pitch-angle to orient radially (i.e $\theta_z' = 90^{\circ}$).
  \subsubsection{Keplerian shear torque: explaining the pitch-angle}
  \label{subsub4}

  While real wakes are transient agglomerations of particles in a shearing disk, we can reproduce the orientation of these structures by introducing the torque on a rigid bar caused by collisions with particles moving past the wake due to the Keplerian shear. To compute this torque, we have to consider the 3-Body Hill's problem. As done in \citet{MORISHIMA} and \citet{yasui} for the case of embedded moonlets, we integrate the Hill equations in 3D to find the colliding trajectories of the ring particles:

\begin{equation}
\begin{split}
  \ddot{x} - 2\dot{y} = \frac{\partial U_H}{\partial x}\\
  \ddot{y} + 2 \dot{x} = \frac{\partial U_H}{\partial x}\\
  \ddot{z} = \frac{\partial U_H}{\partial z}
  \label{3dhill}
\end{split}
\end{equation}

The Hill potential ($U_H$) consists of the tidal potential ($\frac{3}{2}x^2$) plus the potential of a line of mass of finite size, modified to allow for arbitrary wake orientations. The zero-velocity curves for this potential are shown in Figure \ref{zerovel}; once the collisions change the orientation of the wake, the trajectories are recalculated with the updated potential. The particles that are radially inward orbit faster than the CM of the wake, and hence they will tend to align the wake with the azimuthal direction; the radially outward particles have the same effect. To find the colliding trajectories we integrate the Hill equations to then compute the rate of momentum transfer into the self-gravity wake:

  \begin{figure}[t]
    \centering
    \includegraphics[width = 0.7\linewidth]{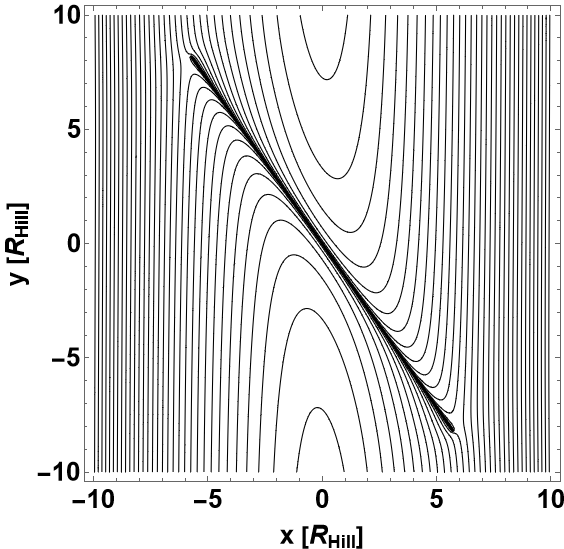}
    \caption{Zero-velocity curves of the Hill potential of a line of mass oriented $45^\circ$ from the azimuthal. The self-gravity wake's potential is modeled as a uniform line of mass of length $L$.}
    \label{zerovel}
  \end{figure}
  
  \begin{equation}
    \tau_{\mathrm{Kep};z'} = 2(1+\epsilon)\int y'\frac{dp_{\perp}}{dt} =2(1+\epsilon)\int^{z_\mathrm{max}}_{z_\mathrm{min}}\int^{\frac{L}{2}}_{0} \rho_s(z)|v_{\perp}|v_{\perp}y'dy'dz'
  \label{torquez}
\end{equation}

\noindent where $\rho_s$ is the space density of the colliding particles and $p_{\perp}$ and $v_{\perp}$ are the momentum and velocity perpendicular to the long axis of the wake, $\epsilon$ is the coefficient of restitution. To numerically compute this integral we create a grid of particles for the incoming impact parameter $b$ and the vertical coordinate $z$, where $\Delta b = \Delta z = 0.01r_H$ where $r_H$ is the Hill radius that was taken to be $\frac{W}{2} = 9  \, \mathrm{m}$ (which is consistent with our value for $\rho_{\mathrm{Roche}}$). The highest and lowest initial $z$ values of particles that collide with the wake are $z_{\mathrm{max}}$ and $z_{\mathrm{min}}$, respectively. We use the vertical density profile $\rho_s(z) = e^{-(\frac{z}{z_0})^2} $ where $z_0$ is the rings half thickness which was taken to be $5 \, \mathrm{m}$ and a temporal resolution of $dt = 0.01 T$ where $T$ is the orbital period at resonance.

We proceed to study the angle at which the torques equilibrate in the case of a single self-gravity wake as a way to test our framework. The total torque in the z-axis can be written as:

  \begin{equation}
    \tau_{z'} = \tau_{\mathrm{tidal}}(\theta_{z'}) + \tau_{\mathrm{Kep};z'}(\theta_{z'},\rho_s, \epsilon)
    \label{equilibrate}
  \end{equation}

  We find that these torques equilibrate stably at $\theta_{z'} = 25^\circ$ when $\rho_s \approx 425,\, 500  \, \mathrm{kg/m^3}$ for $\epsilon = 1, \, 0.5$, respectively. This equilibrium angle is a boundary condition for our problem and any pair ($\epsilon$, $\rho_s$) that achieves it yields the same wake motion inside the wave region. Figure \ref{shear} show how the pitch-angle derived by equilibrating the torques matches the one suggested by UVIS occultations \citep{Colwell2006, Jerousek}, and aligns with the pitch-angles apparent in shear-box simulations conducted with different shear rates \citep{michikoshi, salo2018}. As a reference for these space density ($\rho_s$) values, consider that a preliminary multiwavelenght analysis suggests an average space density of $85  \, \mathrm{kg/m^3}$ in between wakes \citep{Jthesis}; with this density the torques equilibrate at $\theta_{z'}= 50^{\circ}$ for $\epsilon = 1$. The higher value of $425 \, \mathrm{kg/m^3}$ required by the $25^{\circ}$ pitch-angle suggests that wake-to-wake collisions---which are not included in our model---can significantly elevate the average amount of material interacting with a wake within an orbital period.} Note that $\rho_s$ is overestimated by our parallelepiped model for the wake: the lesser inertia of a more realistic ellipsoidal wake would increase the angular acceleration due to collisions while not affecting the tidal angular acceleration. This decreases the required $\rho_s$ by the ratio of the inertia coefficients between the two models, $3/5$. 
  
  \begin{figure}[t]
  \centering
    \includegraphics[width = 0.9\linewidth]{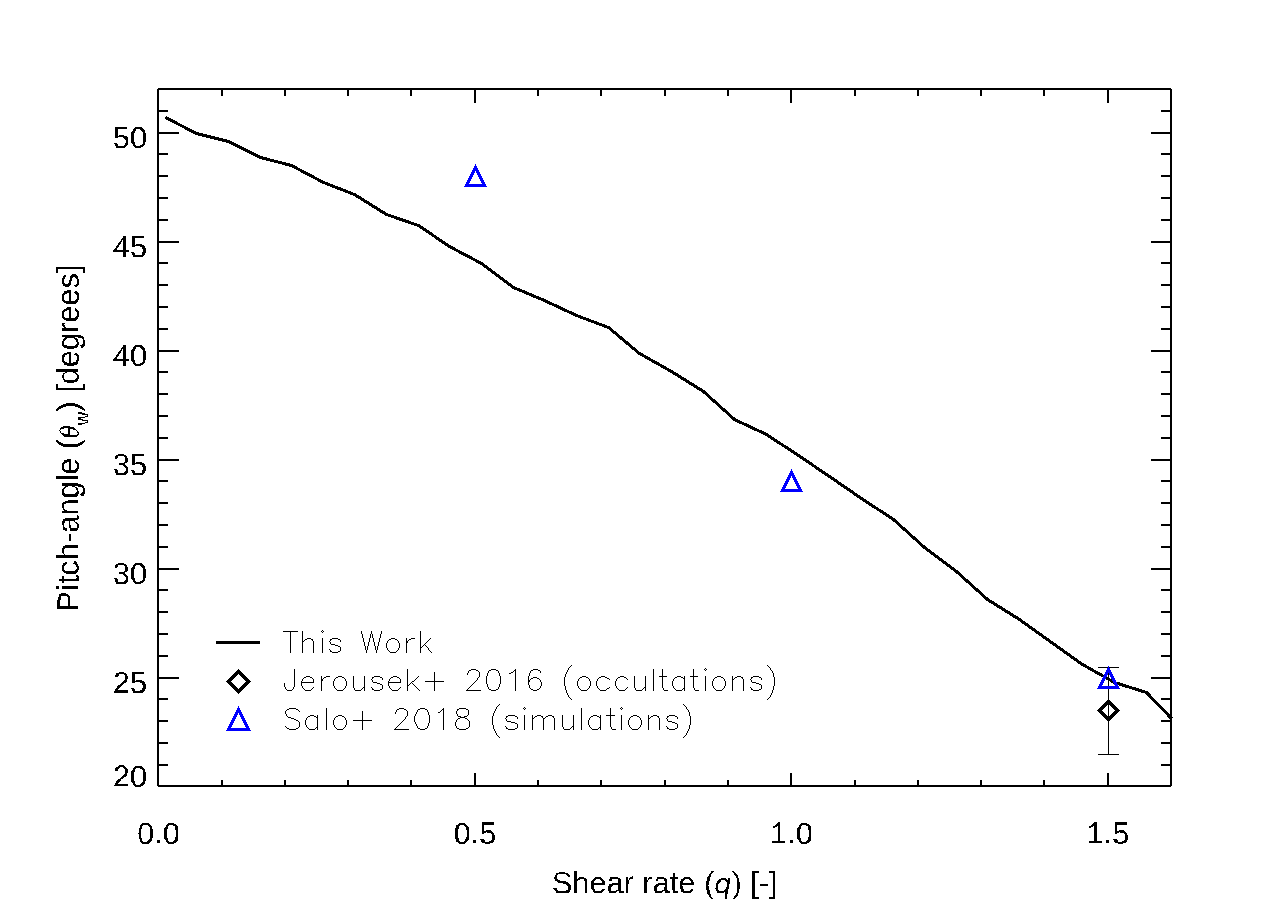}
    \caption{Relationship between the pitch-angle $\theta_w$ (see Figure \ref{wakes}) and the shear rate $q=-\frac{d\ln{\Omega}}{d\ln{x}}$ where $\Omega$ is the mean motion frequency and $x$ the radial coordinate from Saturn. We calibrate our model to equilibrate at $\theta_w = 25^{\circ}$ for $q=1.5$ which is the case for the Mimas 5:3 BW. The blue triangles correspond to equilibrium configurations extracted from a shear-box simulation presented in Figure 16.23 of \citet{salo2018}. The black square represents the values reported in \citet{Jerousek} at the region near the Mimas 5:3 BW (at a Saturncentric distance of $13.2$ Mm) with the error bars representing the range of the values presented in Figure 7 of that work; the upper error bar given by the pitch-angle derived from UVIS occultations. Our rigid-body model matches the shear-box simulations' equilibrium angle within $4^{\circ}$.}
    \label{shear}
  \end{figure}

  The computationally intensive nature of integrating the Hill equations self-consistently forces us to resort to a two-body approximation of the three-body problem. Specifically, a gravitational focusing approximation common in planet formation \citep{Greenberg, Rafikov2004, Armitage}, where the incoming speed of the scattering event can either be the dispersion velocity or the Keplerian velocity depending on which one dominates in the wave region. \citet{cuzzi} already argued that the dispersion velocity dominates over the extent of a particle, but we find that both velocities are similar over the extent of the Hill radius of a wake. Thus, the safest approach for our case is to include 3-body corrections in our two-body approximation. From \citet{Greenberg} and \citet{Armitage} we can get an approximate collision rate for our regime:

  \begin{equation}
    \frac{dM}{dt} = q\Omega \cdot \sigma \cdot f \cdot b \cdot db
    \label{armitage}
 \end{equation}

 \noindent where $f$ is the fraction of particles that enter the Hill sphere and collide, $q$ is the shear rate  $q=-d\ln{\Omega}/d\ln{x}$ ($q=1.5$ for a Keplerian disk), $\Omega$ is the mean motion frequency of the particle, $b$ is the impact parameter and $\sigma$ is the surface mass density of the disk. To get a torque we use two-body dynamics to find the speed and angle of incidence for the incoming particles at the moment of impact. 

 \begin{equation}
   \tau_{\mathrm{Kep};z'} = -2(1+\epsilon)\int^{b_{\mathrm{max}}}_{b_{\mathrm{min}}} \frac{dM}{dt}v_{\theta}(b,\theta_{z'})r_{2B}(b,\theta_{z'})
   \end{equation}

\noindent where $v_{\theta}(b) = b^2*q\Omega/r_{2B}$ and \[r_{2B}(b,\theta_{z'})=\frac{\frac{(b^2q\Omega)^2}{G\rho_{\mathrm{Roche}}WLH}}{1 + e\cos(\theta_{z'} -\varpi)}\] is the two-body orbit equation, where $\varpi$ is the argument of periapsis taken from the azimuthal direction. We have set the true anomaly to the angular position at which the wake is oriented, which is the point in the orbit where the collision takes place (valid for the case of small $\theta_{x'}$ and $\theta_{y'}$). Here we are taking the angular momentum to be conserved about the CM of the wake, so that for a given impact parameter $b$ we can determine a collisional speed $v_\theta$ at a distance $r_{2B}$ from the CM. Under this approximation the trajectories of the colliding particles are hyperbolas and the three-body effects are accounted for by using Hill's equations to determine the limit of the integral $b_{\mathrm{min}}$. This and the length of the long axis ($L$) relative to the Hill radius, which guarantees that particles entering the Hill sphere collide with the wake, allow us to do the following simplification of the \citet{Greenberg} formulation:

\begin{itemize}
\item $f$, the fraction of particles that collide after entering the Hill Sphere, is 1.
\item Tadpole and horseshoe orbits are excluded from the colliding particles by using the three-body code to determine $b_{\mathrm{min}}$. We find that $b_{\mathrm{min}}$ varies from $35  \, \mathrm{m}$ to $43  \, \mathrm{m}$ with the different orientations of the wake (from radially aligned to the azimuthally aligned), with a mean value of $b_{\mathrm{min}}=  38  \, \mathrm{m}$, which we use as a constant $b_{\mathrm{min}}$ for the two-body simulation.
\item We define $b_{\mathrm{max}}$ as the solution of $r_{2B}(b_{\mathrm{max}},\theta_{z'}) = \frac{L}{2}$
\end{itemize}

  The collisional torque then becomes:
  \begin{equation}
    \tau_{\mathrm{Kep};z'} = -2(1+\epsilon)\int^{b_{\mathrm{max}}}_{b_{\mathrm{min}}} \left|v_{\theta}+\omega_{z'}r_{2B}\right|(v_{\theta}+\omega_{z'}r_{2B})r_{2B} \int^{\frac{H}{2}}_{-\frac{H}{2}}\rho_s(z) dz db 
    \label{twobody}
  \end{equation}

  \noindent where we have replaced the incoming speed in Eqn. (\ref{armitage}) ($q\Omega b$) with $|v_\theta|$ and used $\int^{H/2}_{-H/2}\rho_s(z) dz$ for $\sigma$. Additionally, by adding $\omega_{z'}r_{2B}$, we allow for the rotation of the wake to affect the incoming relative speed of the particles. This integral is solved numerically since we have to determine $b_{\mathrm{max}}$ every time the pitch angle $\theta_{z'}$ changes. For a Keplerian shear rate, we find that this formulation equilibrates stably with the tidal force (Eqn. \ref{tidal}) at $\theta_{z'} = 25^{\circ}$ for $\rho_s = 500  \, \mathrm{kg/m^3}$ and $\epsilon = 1$. Therefore Eqn. (\ref{equilibrate}), with the Keplerian shear torque given by Eqn. (\ref{twobody}), meets the boundary condition for the orientation of the wake outside of the wave. We proceed to explore the torques and the motion of the wake's inside the BW region.

  \subsubsection{BW acceleration torque}
\label{subsub1}
The main torque that causes the wakes' rotation in the BW region stems from the difference between the acceleration of the CM and that of a wake particle a radial distance $\Delta x $ from the CM; we write this as $\ddot{z}_{\mathrm{CM}} - \ddot{z}(x)$. Given that all the torques are derived in the frame of the CM of the wake, differences in acceleration will appear as a force. To compute this force we take two time derivatives of Eqn. (\ref{h}), which denotes the vertical position of a ring particle as governed by the interactions in SCL theory; namely, the self-gravity of the bent ring, the gravitational force of Saturn, and a viscous force. Note that the time dependence of Eqn. (\ref{h}) lies entirely in the exponential factor $e^{i\omega' t}$, and, given the $\sim 10 \, \mathrm{km}$ wavelength and the tightly-wound nature of the wave, the change of $z(x,\Theta)$ over a wake can be approximated linearly on $\Delta x$. The difference in vertical acceleration can then be expressed as:
\begin{equation}
  \ddot{z}_{\mathrm{CM}} - \ddot{z}= -\omega'^2({z}_{\mathrm{CM}} - {z}) \approx -\omega'^2 \left. \frac{dz}{dx}\right|_{\mathrm{CM}} \Delta x
  \label{verticalshear}
\end{equation}

Let's consider the rotation about the long axis (the $y'$ axis). We need to find the directional derivative of $\ddot{z}(x)$ in the directions perpendicular to $y'$. Given the flattened nature of self-gravity wakes \citep{Colwell2006, hedman, Jerousek}, variation of forces across the vertical axis, $H$, will be ignored and only variations along $x'$ and $y'$ will be considered. Then for the torque about $y'$ we only need the variation of $\ddot{z}$ along the $x'$ axis. The force then becomes:

\begin{equation}
  F_{z}(x') = -\omega'^2 \left.\frac{dz}{dx}\right|_{\mathrm{CM}}(\hat{x} \cdot \hat{x'}) x' dm
\end{equation}

\noindent where $dm$ is a differential mass a distance $x'$ from the CM. Projecting it onto $\hat{z}'$ and crossing it with the lever arm we get:

\begin{equation}
 \vec{\tau}_{\mathrm{BWacc},y'} = \vec{x}' \times \vec{F_{z'}} =  -2\omega'^2 \left. \frac{\partial z}{\partial x'}\right|_{CM} (\hat{z} \cdot \hat{z}') (\hat{x}' \times \hat{z}') \int^{\frac{W}{2}}_0 \rho_{\mathrm{Roche}}HLx'^2dx'
  \label{torquey}
  \end{equation}

 \noindent where we substituted $dm = \rho_{\mathrm{Roche}} H * L * dx'$ and the directional derivative along $x'$ defined as $\frac{dz}{dx}\Bigr|_{\mathrm{CM}}(\hat{x} \cdot \hat{x}') \equiv \frac{\partial z}{\partial x'}\Bigr|_{CM}$.


  \subsubsection{BW shear torque}
  \label{subsub2}

 The wake's CM vertical speed will differ from that of the neighboring particles and hence there will be a collision which will generate a torque. This is similar to the Keplerian torque, and, for the general case, the vertical and horizontal speeds are added to compute a total collisional torque (see Appendix \ref{AA}). If $\theta_{x'}$ and $\theta_{y'}$ are small, however, these two torques are independent. To model the BW shear torque we consider the shear $\frac{d\dot{z}}{dr}$ and note that the particles a distance $y'$ along the long axis will have a relative vertical velocity of $\frac{d\dot{z}}{dr}(\hat{x} \cdot \hat{y'})$ with respect to the CM. If these particles have a space density $\rho_{s}$ then the amount of mass colliding with the wake at a distance $y'$ in a time $dt$ will be:

  \[M(r)  =  \rho_{s} \left|\frac{\partial \dot{z}}{\partial y'}\Bigr|_{CM} y'dt\right| W dy' \]

  \noindent which causes a momentum transfer of:
    \[dp_{z'}(y') =[M]*[\Delta v_{z'}] = [\rho_{s} \left|\frac{\partial \dot{z}}{\partial y' }\Bigr|_{CM} y'\right| dt W dy']*[(1+\epsilon)\frac{\partial \dot{z}}{\partial y'}\Bigr|_{CM} y']\]

  This momentum transfer corresponds to a force of:

  \begin{equation}
    F_{\mathrm{BWsh};x'}=\frac{dp_{z'}}{dt} = (1+ \epsilon) \rho_{s} \left| \left(\frac{\partial \dot{z}}{\partial y'}\Bigr|_{CM} - \omega_{x'}\right)y'\right|\left(\frac{\ \partial \dot{z}}{\partial y'}\Bigr|_{CM} - \omega_{x'}\right)W y' dy'
    \label{force1}
  \end{equation}

  \noindent where we have included $\omega_{x'}$, the angular velocity about the $x'$-axis, to account for the change in the relative incoming speed that occurs when the wake rotates. The torque associated with this momentum transfer is:
  \begin{equation}
    \tau_{\mathrm{BWsh};x'} = 2\int^{\frac{L}{2}}_0{y' \frac{dp_{z'}}{dt}} = \frac{1+\epsilon}{2} \rho_s W  \left|\frac{\partial \dot{z}}{\partial y'}\Bigr|_{CM} - \omega_{x'}\right|\left(\frac{\partial \dot{z}}{\partial y'}\Bigr|_{CM}-\omega_{x'} \right) \left(\frac{L}{2}\right)^4
  \label{taucolx}
\end{equation}
 \noindent where we have taken the wake to be much more massive than the incoming ring particles so that the CM of the collision is the CM of the wake. Doing the same computation for $\tau_{\mathrm{vert};y'}$, we can arrive at the total torques about the $x'$ and $y'$ axis, now written as a function the angle $\theta_{z'}$ and radial distance $x$. For small angles $\theta_{x'}$ and $\theta_{y'}$ we have:

  \begin{equation}
    \tau_{x'} \approx -\frac{2}{3}\rho_{\mathrm{Roche}}HW  \frac{d \ddot{z}}{dx}\Biggr|_{CM}\sin{\theta_{z'}}\left(\frac{L}{2}\right)^3 + \rho_s W \frac{1+\epsilon}{2} \left|\frac{d \dot{z}}{dx}\Biggr|_{CM}\sin{\theta_{z'}} - \omega_{x'}\right|\left(\frac{d \dot{z}}{dx}\Biggr|_{CM}\sin{\theta_{z'}}-\omega_{x'}\right) \left(\frac{L}{2}\right)^4
    \label{torquex2}
  \end{equation}
  
  \begin{equation}
    \tau_{y'} \approx \frac{2}{3} \rho_{\mathrm{Roche}}HL \frac{d\ddot{z}}{dx}\Biggr|_{CM}\cos\theta_{z'} \left(\frac{W}{2}\right)^3 + \rho_s L \frac{1+\epsilon}{2} \left|\frac{d \dot{z}}{dx}\Biggr|_{CM}\cos{\theta_{z'}} + \omega_{y'}\right|\left(\frac{d \dot{z}}{dx}\Biggr|_{CM}\cos{\theta_{z'}}+\omega_{y'}\right) \left(\frac{W}{2}\right)^4
    \label{torquey2}
\end{equation}

\noindent where, in this approximation, $\theta_{z'}$ is the pitch-angle of the wake.

\subsubsection{Motion of an isolated self-gravity wake}
\label{subsub5}

Real wakes consist of many transient objects that can gravitationally interact. However, to understand the underlying dynamics within a BW, it is worth considering the simpler motion of an isolated wake subjected to the torques computed thus far; later, we will include the effects of surrounding wakes. Unlike the case of the flat ring, inside the BW the wakes are continuously perturbed via out-of-the-plane torques; thus, there is no torque equilibrium and we must integrate the equations of rigid motion (Eqns. \ref{eulerx}, \ref{eulery}, and \ref{eulerz}) These equations are numerically integrated with a midpoint method \citep{numerical} (differential rotation is commutative, so the order of rotation does not affect the integration).

Note that the inertial terms
 \begin{equation}
    \tau_{\mathrm{inertia}_k}=(I_{i} - I_{j})\omega_i\omega_j
    \label{inertia}
  \end{equation}

 \noindent (where $i \ne j \ne k$  and $i,j,k=x',y',z'$) will prove important in understanding the wake's rotation since they couple the rotations about the different axis.

\begin{figure}[t]
  \centering
    \includegraphics[width = 0.9\linewidth]{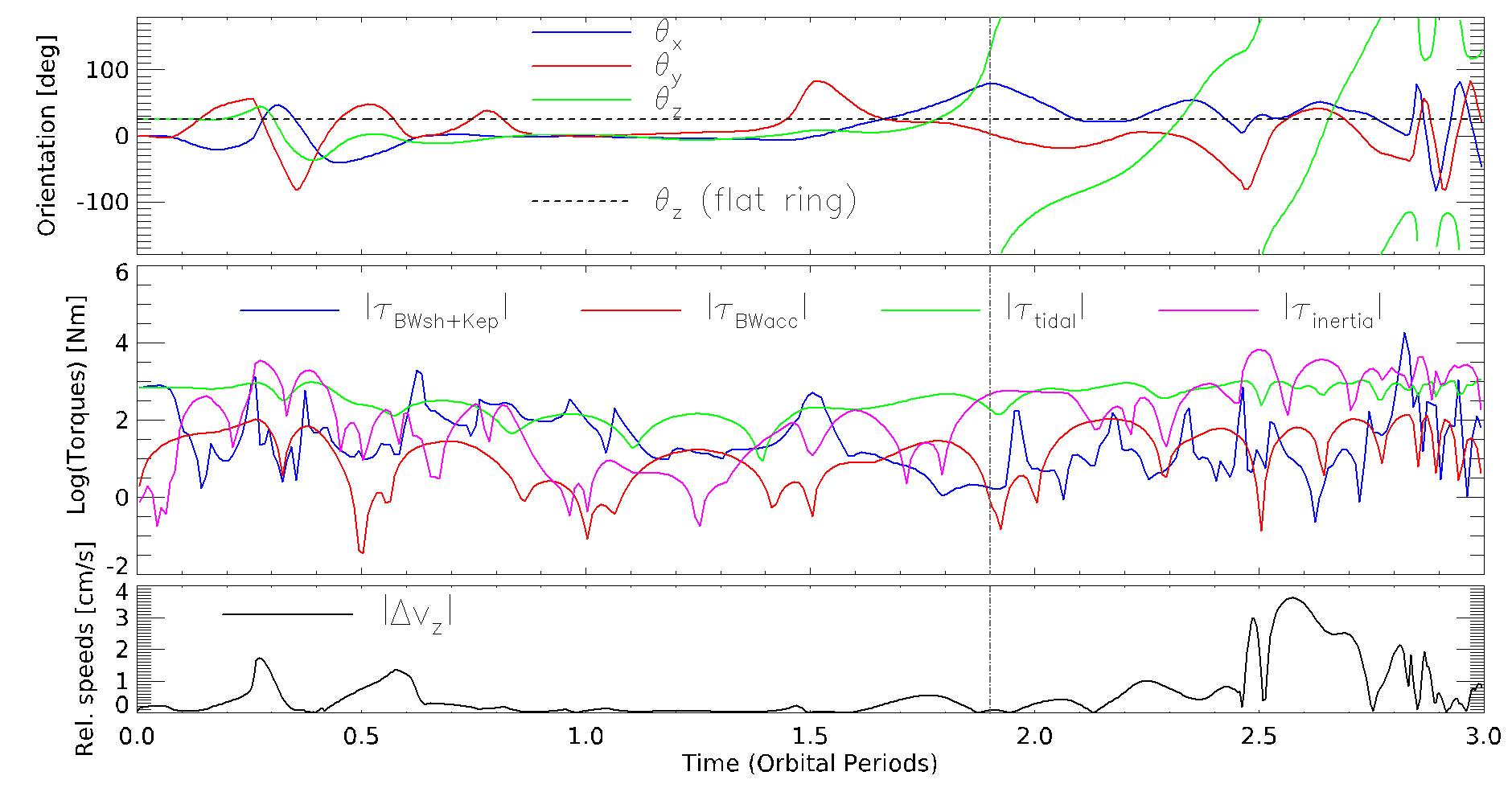}
    \caption{Angular motion (top panel), torques (middle panel), and speed relative to the ring (bottom panel) of a uniform density self-gravity wake satellite evolved integrating the equations of rigid motion (Eqns. \ref{eulerx}, \ref{eulery} and \ref{eulerz}) and the 3D Hill equations (Eqn. \ref{3dhill}).$\theta_z=0$ when the long axis ($y'$), projected onto the plane, points to the azimuthal, $\theta_x=0$ and $\theta_y=0$ when the $x'$ and $y'$ axes, respectively, lie in the rings' plane and $\theta_x, \theta_y90^{\circ}$ when they point in the vertical direction. The wake has uniform density and dimensions $W=18 \, \mathrm{m}$, $L=232 \, \mathrm{m}$ and $H=4 \, \mathrm{m}$, and the coefficient of restitution $\epsilon$ and the space density $\rho_s$ of the ring are set to balance the torques outside of the wave at a pitch-angle is $25^{\circ}$ (see Figure \ref{wakes} and related discussion). The maximum slope is set to $14^{\circ}$ and the slope is $0$ at $t=0$, and varies harmonically with the forcing frequency $\omega'$ (see Eqn. \ref{omega}). For comparison, we also show the time evolution of the orientation of a wake on a flat ring (black dashed line). Top panel: the wake inclines back and forth as the wave passes through, and its long axis spins around in the plane until it reaches a steady state where the wake oscillates about the azimuthal direction. Marked with a dashed gray line is the point where the motion becomes too extreme, given that $\theta_x = 80^{\circ}$, and given that the wake points completely in the radial direction. Middle panel: the torques associated with the displayed motion. The inertial torque becomes important, and the BW acceleration torque is on average the smallest one by two orders of magnitude. Bottom panel: z-component of the relative velocities between the wake and a particle colliding with the edge of the wake.}
  \label{nowakepotential}
\end{figure}

 In Figure \ref{nowakepotential} we describe the rotation of the wakes in the BW, for a maximum slope of $|dz/dx|_\mathrm{max}=14^{\circ}$. The top panel shows the change in the wake's orientation, the middle panel shows the magnitude of the torques, and the bottom panel shows the vertical impact speeds. To calculate Figure \ref{nowakepotential} we relaxed the approximation that $\theta_{x'}$ and $\theta_{y'}$ are small, which amounts to computing all 4 torques symmetrically for the 3 axes. The full form of the torques can be found in Appendix \ref{AA}.

 In order to describe the orientation of the wake in inertial space, we have introduced the angles $(\theta_x,\theta_y,\theta_z)$ (upper panel), which correspond to the wakes' `pitch', `roll', and `yaw', respectively. $\theta_x=\cos^{-1}{(\hat{x}' \cdot \hat{z})}$ and $\theta_y=\cos^{-1}{(\hat{y}' \cdot \hat{z})}$ increase as the $x'$ and $y'$ axes, respectively, point towards the vertical direction. $\theta_z=\tan^{-1}\left(-\hat{y}' \cdot \hat{x}/\hat{y}' \cdot \hat{y}\right)$ is $0$ when the $y'$-axis, projected onto the ring's plane, points at the azimuthal and it increases counterclockwise. In the specific configurations of the wakes in Figure \ref{frames}, these angles are equivalent to their corresponding primed counterparts. Moreover, if $\theta_{x'}$ and $\theta_{y'}$ are small, then $\theta_{z'}\approx \theta_{z}$.

  In the top panel of Figure \ref{nowakepotential} we show the motion of a wake when the slope of the wave is constant and zero (black line). We find that the wake equilibrates stably at an angle $\theta_z=25^{\circ}$. However, when the slope is changing, the orientation of the wake becomes significantly more complex (colored lines in the top panel). The sole difference causing this divergence in behavior is the introduction of a slope via the BW acceleration and shear torques---even if this slope is relatively small ($|dz/dx|_{\mathrm{max}}=\tan{14^{\circ}}$). The dependence on the small slope and the forcing frequency ($\omega'$) makes the BW torques (on average) two orders of magnitude smaller than the Keplerian and the tidal torques (middle panel), but the inclinations they cause provoke a non-linear behavior in the angular variables $(\theta_{x'} $,$\theta_{y'}$, $\theta_{z'})$ which are strongly coupled. The coupling mainly comes from the inertial term.

  Since the BW torques are small, the inclination of the wake comes from the action of this inertial term Eqn. \ref{inertia}), but the inertial term is non-zero only due to the presence of the BW torques. Therefore, the smallest torques in magnitude, namely the ones introduced by the BW, have the biggest effect on the dynamics. The wakes no longer stay on the plane of the ring simply oscillating about an equilibrium orientation; rather, they spin full circles and reach inclinations of $80^{\circ}$ from the equatorial plane for the case of a maximum slope of $14^{\circ}$ (this occurs at the vertical dashed line).

The bottom panel of Figure \ref{nowakepotential} shows the relative vertical speed between the wake and the ring particles. These are computed as

\begin{equation}
  v_{\mathrm{rel};z'}=[\frac{L}{2}(-\omega_{x'} + \frac{d\dot{z}}{dx}) + \frac{W}{2}*(\omega_{y'} + \frac{d{\dot{z}}{dx}})](\hat{z'} \cdot \hat{z}) + \frac{L}{2}\omega_{x'}(\hat{x}' \cdot \hat{z})
\end{equation}

\noindent where the dominant terms are the projections of the rotational speeds onto the $z$-axis. The relative vertical velocities within the wake and the ring particles associated with this motion are in the $\mathrm{cm/s}$ range (bottom panel of Figure \ref{nowakepotential}). Note, however, that we are still to account for the gravitational interactions between self-gravity wakes, which, as we will see, can have a stabilizing effect on the pitch-angle of the wakes.
 
\subsubsection{Torque due to self-gravity wakes}
\label{subsub6}

  For the gravitational force that self-gravity wakes exert on each other we consider the case of a wake with variable orientation embedded on a field self-gravity wakes oriented at a constant $\theta_{z'}=\theta_{w}$. Taking this field as a set of density plane waves, we can use Pringle and Lyden-Bell equations for the torque of a spiral arm on a galaxy \citep{PandLB}, which have been applied to self-gravity wakes in proto-lunar disks in \citet{2001ApJ}. The potential of the spiral arms is given in \citet{BandT} in terms of Saturncentric distance $r$ and longitude $\Theta$ as:

  \begin{equation}
    \Phi' = \frac{2 \pi G}{|k|} \sigma_w Re[e^{i(m_w \Theta + k_r r)}]
      \label{arms}
    \end{equation}

    \noindent where $k=\frac{2 \pi}{\lambda_{T}}$ and $m_w$ and $k_r$ are the azimuthal and radial wavenumbers, and $\sigma_w$ is the wake surface density minus the mean surface density of the ring.

    Following the appendix in \citet{2001ApJ}, we relate the wavenumbers by $m_w = k r \sin\theta_{w}$ and $k_r = k\cos\theta_{w}$, where $\theta_{w}=25^{\circ}$ is the pitch-angle. To compute $\sigma_w$  we will approximate the wake surface density to be $\rho_{\mathrm{Roche}}\int^\infty_{-\infty} e^{-(\frac{z}{H/2})^2}dz  \rho_{\mathrm{Roche}}\frac{\sqrt{\pi}}{2}H $, noting that the mean surface density $\sigma$ for the A ring region is about $400 \, \mathrm{kg/m^2}$ \citep{Tiscareno} and has a negligible effect on $\sigma_w$.  We will now expand the potential (Eqn. \ref{arms}) about the center of a wake located at $(r_0, \Theta_0=y_0/r_0)$ where $y_0 = 0$, similar to what is done in \citet{cook}. The potential then becomes:

    \begin{equation}
      \Phi' = \frac{2 \pi G}{|k|} (\rho_{\mathrm{Roche}}\frac{\sqrt{\pi}}{2}H)(1 - k^2(\cos{\theta_w}(r-r_0) + \sin{\theta_w} y)^2)
    \end{equation}

To write the potential in terms of the orientation of the wake $\theta_{z'}$ (in the approximation where $\theta_{x'}$ and $\theta_{y'}$ are small), we consider the distance about the long axis, $y'$, and write $r-r_0 = -y'\sin(\theta_{z'})$ and $y = y'\cos(\theta_{z'}$)

This yields the potential

\[ \Phi' = \frac{2 \pi G}{|k|}(\rho_{\mathrm{Roche}}\frac{\sqrt{\pi}}{2}H)[1 - k^2y'^2\cos^2{(\theta_{z'}-\theta_w)}] \]

We now apply the $-\frac{1}{y'}\frac{\partial}{\partial \theta_{z'}}$ operator to the potential and multiply by the differential mass times the lever arm ($dm \cdot y' =\rho_{\mathrm{Roche}}HWdy' \cdot y'$) to get the torque at distance $y'$ from the center of the wake; we then proceed to integrate over $y'$ and get:
  
\begin{equation}
  \tau_{\mathrm{wake};z'} = \frac{4 \pi^2}{\lambda_T} G \rho_{\mathrm{Roche}}\frac{\sqrt{\pi}}{2}H^2 W \frac{L^3}{24}*\sin{2(\theta_{w}-\theta_{z'})}
  \label{waketorque}
\end{equation}

  \noindent which changes the equation of the torque about the z-axis to:

  \begin{equation}
    \tau_{z'} = \tau_{\mathrm{tidal};z'}(\theta_{z'}) + \tau_{\mathrm{Kep};z'}(\theta_{z'},\rho_s) + \tau_{\mathrm{wake};z}(\theta_{z'})
    \label{torquez2}
  \end{equation}

  The torques in the z-axis now equilibrate at an angle of $\theta_{z'} = 25^{\circ}$ for $\rho_s = 390  \, \mathrm{kg/m^3}$ for the full Hill equations computation and $\epsilon = 1$. The equilibrium orientation of the wake is now less sensitive to the surrounding space density since the wake potential creates a minimum when wakes align (note that Eqn. \ref{waketorque} vanishes when $\theta_{z'}=\theta_{w}$). This may explain why the pitch-angle of the wakes is nearly uniform over the rings \citep{Jerousek} even if the density and the shear rate vary between the A and the B rings. We are assuming the coherent structure of $\theta_w =25^{\circ}$ is kept within the wave, but the wakes may not have a coherent orientation or well-defined wavenumbers, which is suggested by the reduction of the number of coherent gap structure in UVIS occultations to $25\%$\ the outside value \citep{morgan} and by our equations of motion which predict that the wakes can spin up or oscillate about the azimuthal; in which case Eqn. (\ref{waketorque}) would be modified by the wakes that are not part of the plane-wave field. While we lack an expression for the case of neighboring wakes with multiple orientations, we will heuristically extend the case of a single misaligned wake embedded in a wake field with orientation $\theta_w$, to the case of a misaligned wake in the BW region embedded in the oriented wake field of the A-ring. The total wake potential acting on a wake is then given by Eqn. (\ref{arms}).

  \subsubsection{Motion of an embedded self-gravity wake}
  \label{subsub7}

\begin{figure}[t]
  \centering
    \includegraphics[width = 0.9\linewidth]{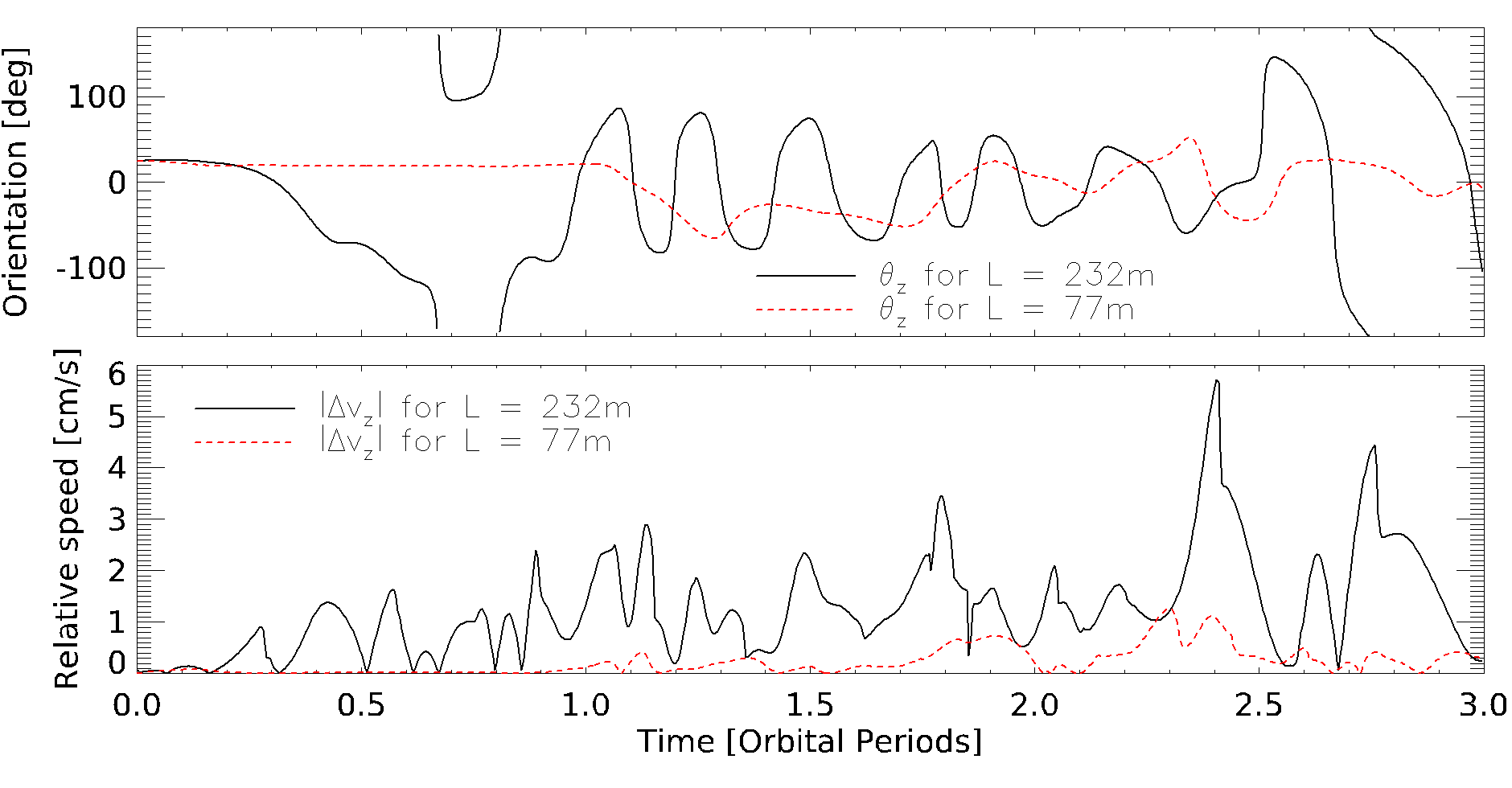}
    \caption{Time evolution of the average velocities of particle collisions with self-gravity wakes within the bending wave; motion evolved as in Figure \ref{nowakepotential} with the addition of the wake potential. The maximum slope is set to $14^{\circ}$ and $t=0$ when the slope is $0$; the slope varies harmonically with the vertical frequency.  Adding the wake potential increases the maximum relative velocity between the wakes and ring particles. For the case of $L=232 \, \mathrm{m}$ we see vertical velocities well within the $\mathrm{cm/s}$ range, while the $L=77 \, \mathrm{m}$ case peaks at about $1 \, \mathrm{cm/s}$.}
    \label{wakepotential}
  \end{figure}

  The relative velocity and orientation of a self-gravity wake embedded on a field of self-gravity wakes with fixed orientation $\theta_w$, evolved under Eqns. (\ref{eulerx}), (\ref{eulery}), and (\ref{eulerz}), is plotted in Figure \ref{wakepotential}.
  
We see a similar non-linear behavior as the isolated wake (Figure \ref{nowakepotential}) resulting in $\mathrm{cm/s}$ collisional speeds (Figure \ref{wakepotential}), but the values of $\theta_{z}$ are more conservative---oscillating about $\theta_{z} = 0$ and staying below $90^{\circ}$ for most of the motion. The restoring force to the equilibrium angle has now increased, and this makes the vertical collisional speeds --- which are coupled to $\theta_{z'}$ by the inertial term (Eqn. \ref{inertia}) --- increase as the wake swings back into position.

The extreme values for the orientation angles, particularly $\theta_z$ which at times has the long axis pointing in the radial direction, make us skeptical that the rigidity assumption holds throughout all of the motion. To remedy this we also computed the motion of a wake with long axis $L=77  \, \mathrm{m}$, a third of the usual length. Note that we are not suggesting that self-gravity wakes will shed two thirds of their mass (that would create a much bigger haze signal than the one observed), rather we are verifying that even if a wake were to shed the regolith near its edges while rotating, it will still maintain collisional speeds in the $\mathrm{cm/s}$ range. Hence, the process is not self-limiting, nor does it necessitate a wake of long axis $L=232 \, \mathrm{m}$.

The rapid change of $\theta_z$ for the $L=232 \, \mathrm{m}$ case also makes us consider the possibility of the release of material due to rotation (e.i rotational disruption). Note that in this scenario, the launched regolith itself will exit the wave at $ \mathrm{cm/s}$ speeds, hence contributing to the haze. In the next section we will show that both, the rotational disruption and collisional ejecta scenarios, create a haze with the same observable features.

  \subsection{Haze predicted properties}
  \label{sec:23}

  The relative velocity curves in Figures \ref{nowakepotential} and \ref{wakepotential} are of the order of $1 \, \mathrm{cm/s}$. Microgravity experiments have shown that at $\mathrm{cm/s}$ velocities we begin to see ejecta in particle collisions with simulant regolith aggregates \citep{primed}. The $\mathrm{cm/s}$ range is also important because the characteristic vertical speeds of the Mimas 5:3 BW are in the $\mathrm{cm/s}$ range: a $\mathrm{cm/s}$ change in velocity is necessary to have particles reset their motion into trajectories that are no longer in phase with the wave.

  If these collisional speeds increase with slope amplitude, the amount of released material would change with radial position. Figure \ref{linear} shows precisely this: the peaks of the relative speed over a period have a monotonically increasing trend with the amplitude of the slope of the wave. To determine how the amount of material changes due to this increase in the velocities, we apply a square-root regression to Figure \ref{linear} and find that:

      \begin{figure}[h!]
  \centering
    \includegraphics[width = 1\linewidth]{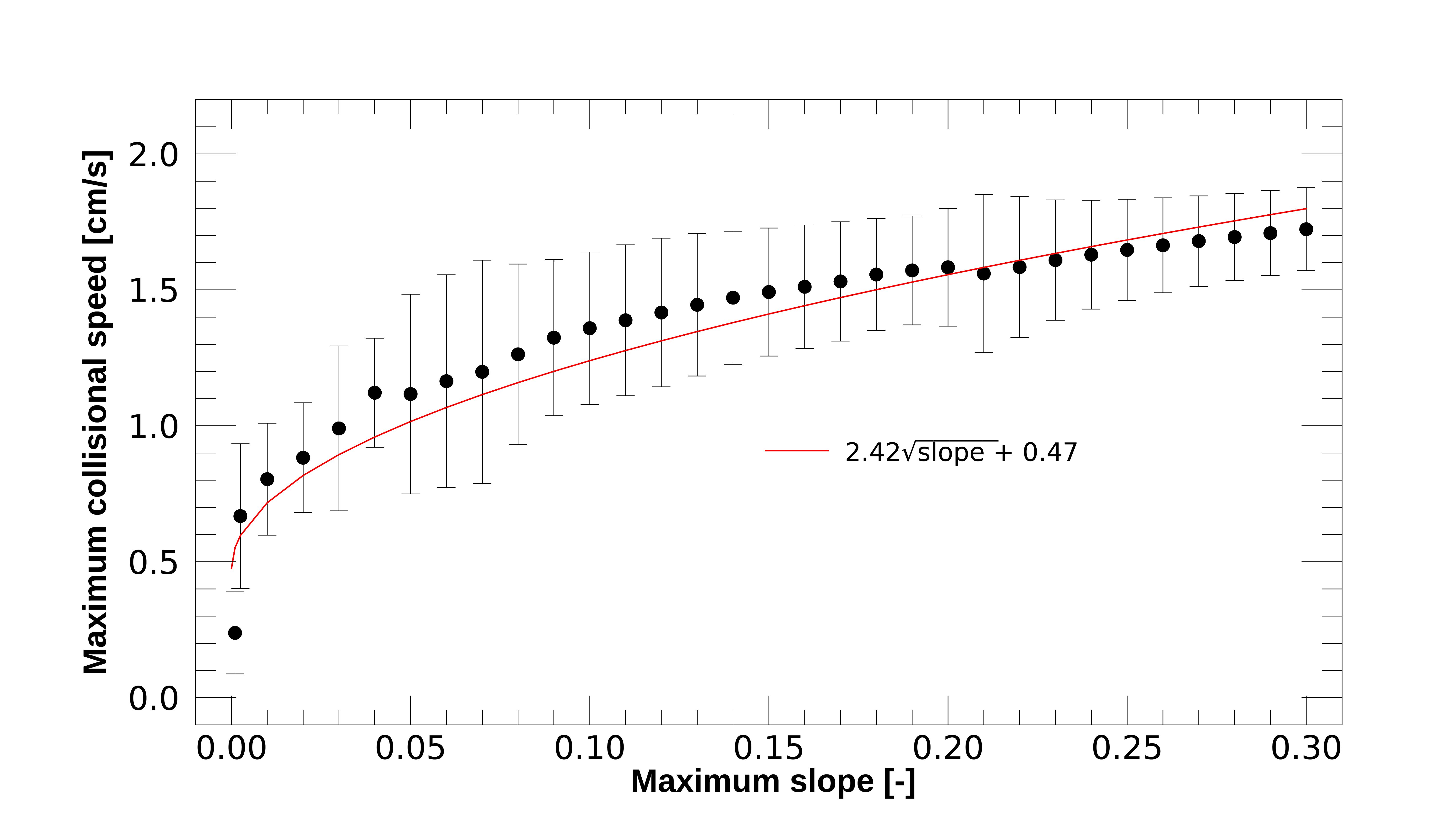}
    \caption{Average maximum relative speed of self-gravity wakes for different slopes. The average is taken over a range of initial conditions to show that the increase of collisional speed with the slope is robust to initial conditions. At each value of the slope considered, we start the wake out-of-phase (at an angle with the ring) by $0$, $\pm10$, $\pm20$, and $\pm30$ degrees at $15$ equidistant points in the wave's vertical motion and run the simulation for 1 period. Therefore, each point is an ensemble average of $105$ runs, where the error bars represent the $1\sigma$ of the distributions. We see a clear trend: the bigger the slope amplitude of the BW the bigger the velocity of the collisions between the wakes and the ring particles. We model this increase with a square-root law.}
    \label{linear}
  \end{figure}

  \begin{equation}
    v_{\mathrm{rel};z} = a\sqrt{\left| \frac{dz}{dx} \right|_{\mathrm{max}}} + w
    \label{vimpact}
  \end{equation}


  \noindent where $a = 2.42 \, \mathrm{cm/s}$ and $w = 0.47 \, \mathrm{cm/s}.$

  Microgravity experiments also show the existence of a speed threshold before which the collisions do not lift ejecta \citep{primed}. Once this minimum threshold has been passed, numerical simulation of collisions in free space \citep{sarah} and drop tower experiments \citep{ejecta} suggest that the amount of ejected regolith increases linearly with the specific kinetic energy of the impactor; we represent this by the equation:

    \begin{equation}
     N_{\mathrm{haze}} \propto \mathcal{H} (v_\mathrm{impact}-g) \cdot (v_{\mathrm{impact}}-g)^2
    \label{n_haze}
  \end{equation}

  \noindent where $N_{\mathrm{haze}}$ is the number of haze particles liberated per wake, $g$ is the impact speed threshold, and $\mathcal{H}$ is the Heaviside function whose value is $0$ if the argument is negative, $1$ otherwise.

  We do not know where this threshold is for the case of wakes in the rings, but it must be higher than the dispersion velocity ($\sim \Omega z_0 = 0.13 \, \mathrm{cm/s}$), since this does not generate sub-mm ejecta outside of the BW (Table \ref{uvisvims}). From now on we heuristically assume that $g = w$ so that $N$ is proportional to the slope amplitude. Misestimating the value for $g$ by a factor of $2$ would change the quantity of haze produced at low slopes ($<$$0.03$) by a factor of $\sim 4$; however, more than $90\%$ of the wave lies in the high-speed regime where the square-root trend works well. Given this value of $g$, and assuming a set number of wakes per area, the amount of haze particles per area ($\Sigma_{\mathrm{haze}}$) has the proportionality of:

  \begin{equation}
    \Sigma_{\mathrm{haze}} \propto \left| \frac{dz}{dx} \right|_{\mathrm{max}}
    \label{SIGMA}
  \end{equation}

  Note that the number density of the haze, $n_{\mathrm{haze}}$, is given by $\frac{\Sigma_{\mathrm{haze}}(r)}{D(r)}$ where $D(r)$ is the vertical thickness of the haze. Combining Eqns. (\ref{vimpact}) and (\ref{SIGMA}) and the equation for the differential optical depth $d\tau_{\mathrm{haze}}(r) = n_{\mathrm{haze}}\bar{\sigma}_{\mathrm{haze}}dl$, where $\bar{\sigma}_{\mathrm{haze}}$ is the typical cross-section of particles in the haze and $dl$ is a differential path-length through the haze, we arrive at the relation:

  \begin{equation}
    d\tau_{\mathrm{haze}}(r) = \frac{\beta}{D}  \left|\frac{dz}{dx}\right|_{\mathrm{max}} dl
    \label{eureka}
  \end{equation}

  \noindent where $\beta=\Sigma_0 \bar{\sigma}_{\mathrm{haze}}$ is the normal optical depth of the haze when the slope is unity, and a free parameter in our model; $\Sigma_0$ is the surface density generated by a slope of unity.

  For the case of rotational disruption, we take the number of released haze particles per wake to be proportional to the centrifugal force due to the rotation: $N_{\mathrm{haze}} \propto \omega_{x'}^2r_p$ where $\omega_{x'}$ is the angular speed and $r_p$ is the radial distance from the CM. Note that the angular speed and $v_{\mathrm{rel};z}$ are related by $\omega_{x'} \approx v_{\mathrm{rel};z}/r_p$; combining this with Eqn. (\ref{vimpact}), we again arrive at Eqn. (\ref{eureka}). Therefore, given our rigid model for self-gravity wakes, we predict that in the BW region the extra optical depth must be proportional to the maximum slope of the wave. Note that $\beta$ is a dimensionless absorption per slope coefficient related to the ejecta-generating efficiency of rotation or collisions.
  
\begin{figure}[h!]
  \centering
    \includegraphics[width=\linewidth]{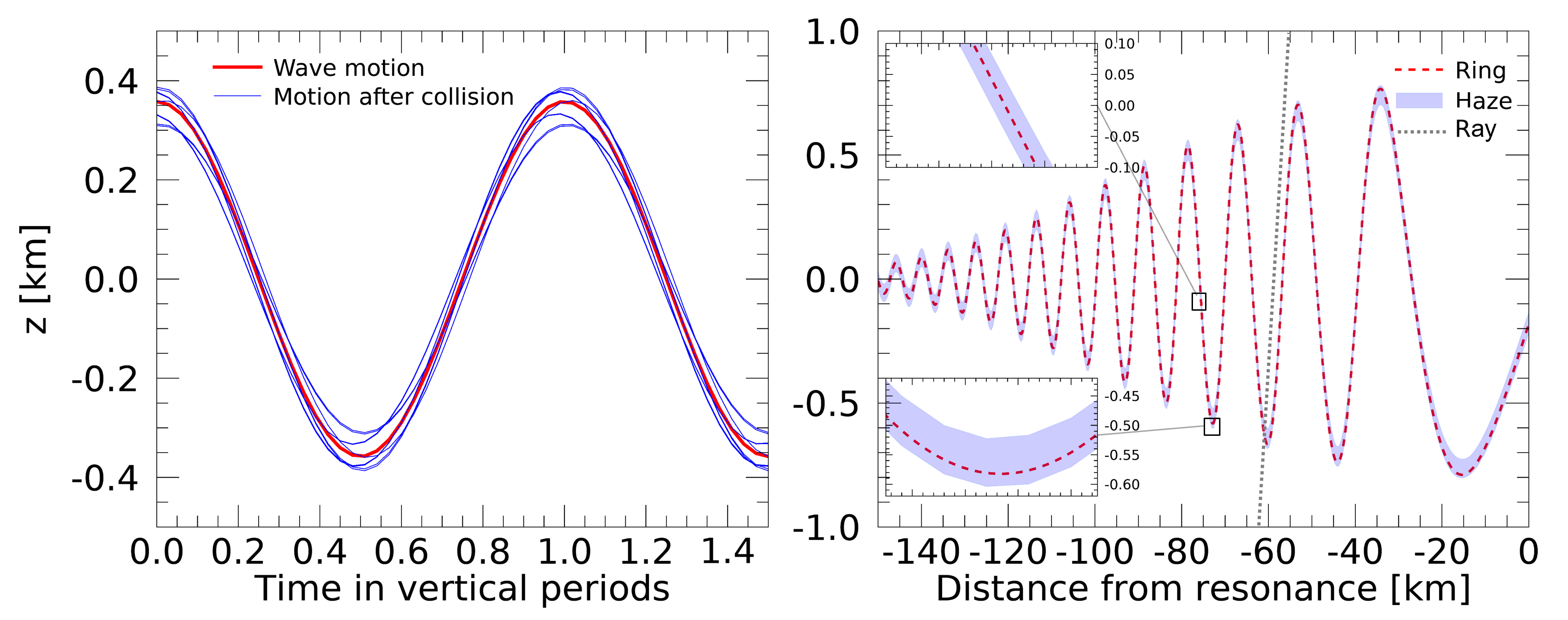}
  \caption{Left panel: trajectories in time of the vertical position of haze particles (blue) and the oscillating ring (red) at $-100 \, \mathrm{km}$ from resonance. Here the trajectories of haze particles start every phase interval of $\frac{\pi}{3}$ during the up-and-down motion (so we are modeling 6 collisions per period). The ring has a continuous envelope of particles during the period. Right panel: the envelope created in space by the trajectories of particles post-collision using $\epsilon = 0.5$. By looking at the lower left insert panel one can see the slight variation of the vertical thickness present in the haze profile; the thickness of the haze changes from $113$ to $134$ m in the overall plot, and the variations in the zoom-in panel are only of $4 \, \mathrm{m}$ (less than a thick mark). A light ray has been plotted in dashed gray to visualize the geometry of the occultations.}
  \label{haze}
\end{figure}

  To construct the geometry of the haze we consider the trajectories of the particles after the collision with the self-gravity wake. Given that in Figure \ref{wakepotential} $>$$1 \, \mathrm{cm/s}$ collisional speeds are achieved at many points during a period, we consider that the collisions with the self-gravity wakes are equally likely to occur at any point in the vertical motion of the particles. Moreover, we consider the imparted velocity to be in equal proportions radial and vertical with a magnitude of $a\sqrt{|\frac{dz}{dx}|_{\mathrm{max}}}+w$ (where $a$ and $w$ are defined in Eqn. \ref{vimpact}). The subsequent trajectories can be easily computed by taking the CM of the collision to be the center of the self-gravity wake. Let $V_{\mathrm{CM}}$ be the velocity of the CM, then the outgoing speed will be given by

  \begin{equation}
    v_0 = V_{\mathrm{CM}}(1 + \epsilon) \pm \epsilon(a \sqrt{\left|\frac{dz}{dx}\right|_{\mathrm{max}}} + w)
    \label{kick}
  \end{equation}

 \noindent the case of $\epsilon=1$ representing the outgoing speed for the rotational disruption scenario. Relative to a particle in the wave the ejected particles move vertically with a Doppler shifted frequency of $\mu_V \pm k_{b}v_{0_x}$, where $\mu_V$ is the vertical frequency of the particle and $k_{b}$ is the wavenumber of the BW. The subsequent trajectories are shown in the left panel of Figure \ref{haze}. In the right panel of Figure \ref{haze} we show the envelope these trajectories form over the radial extent of the wave; note that the thickness of the haze varies radially and is greater at the peaks and troughs than at the equatorial plane. Moreover, the haze's thickness varies with $\epsilon$, and we find it to range from ${90 - 190} \, \mathrm{m}$ for $\epsilon = {0.1 - 1}$.

  While there's phase dependence in the haze's geometry (Figure \ref{haze}), the dynamics predict the same number of particles at any given phase for a given radial location $x$. Hence, the change in the optical depth profile of a normal occultation can only come from the change of scattered particles in the collisions, which is related to the change in the slope amplitude with $x$. We then predict that the phase should not appear in the haze signal, which is a desired feature of the model, given that the observed signal of $B_{\mathrm{eff}} \cong 90^{\circ}$ occultations show no correlation with the phase of the wave.

\section{Transmission Model and Data Reduction}
\label{sec:3}

The transmission of light through the rings is modeled by the radiative transfer equation simplified by the fact that the opacity $\kappa$ of the ring particles is insensitive to wavelength. Moreover, we are working exclusively in the  $110 - 190 \ \mathrm{nm}$ spectral bandpass of the High Speed Photometer (HSP) within UVIS. Given the lower limit in the $100 \, \mathrm{\mu m}$ range for the particle size in the A-ring \citep{HARBISON}, there is no scattering; thus, we are firmly within the geometrical optics limit. As a consequence of this, the intensity $I$ measured by UVIS is simply:

\begin{equation}
  I(r) = I_0e^{-\frac{\tau_n}{\mu}} + b
  \label{I}
\end{equation}

\noindent where $I_0$ is the unocculted star signal, $b$ is background light entering the detector that does not come from the star, $\tau_n$ is the normal optical depth, and $\mu = \sin(B)$ where $B$ is the angle between the ring plane and the light ray. Or, in terms of the transparency $T$

\begin{equation}
  T = \frac{I - b}{I_0}
  \label{obs}
\end{equation}

We will extract the observed optical depth by using:

\begin{equation}
  \ln{\frac{I(r) - b}{I_0}} = - \frac{\tau_n}{\mu}
\end{equation}

In order to isolate the star signal we determine the background signal ($b$) and the unocculted star signal ($I_0$) following the methods of \citet{Colwell2010}. It is known that the detector becomes more sensitive over time when the optical depth is close to $0$ \citep{Colwell2010}, which introduces a systematic uncertainty on our measurements of $I_0$ that we are not considering here. Since $I_0$ refers to the star, we focus on the properties of the rings themselves by working with the transparency ($T$) and the optical depth ($\tau$).

To the $217$ occultations of the BW region we apply a signal-to-noise filter, or conversely a $\tau_{\mathrm{max}}$ filter \citep{Colwell2010} of $\tau_{\mathrm{max}}>1.5$ so that the variations in $\tau$ due to the BW are distinguishable; this reduces the set of occultations to $130$ (see Appendix \ref{AB} for the complete list). From this we select $60$ occultations with the following criteria:

First, we maximize the range of geometries probed by our study without oversampling any particular geometry. Six subsets are created where $B$ can be low angle, $0^{\circ}$ to $30^{\circ}$, intermediate angle, $30^{\circ}$ to $60^{\circ}$, or normal, $60^{\circ}$ to $90^{\circ}$. On the other hand, $\phi$ can be radial ($|\cos{\phi}| > \frac{1}{2}$) or azimuthal ($|\cos{\phi}| < \frac{1}{2}$). We use $|\cos{\phi}|$ to partition the data because it is the physically relevant quantity that appears in all the models that will be tested. The smallest subset, being the low $B$ radial set, has a total of $10$ occultations in the entire dataset, so all other subsets must also contain $10$ occultations to not oversample any geometry, yielding $60$ occultations in total.

Secondly, in the subsets with more than $10$ occultations we select the highest $\tau_{\mathrm{max}}$ occultation for each star, and if there are fewer than $10$ unique stars we select the next highest $\tau_{\mathrm{max}}$ for each star until $10$ are reached. The final dataset is shown in Table \ref{table2}.

Additional critical geometric parameters consist of the longitude of the occultation in the rings relative to the moon Mimas, the vertical position of Mimas, and the vertical velocity of Mimas. These parameters were determined using the Navigation and Ancillary Information Facility (NAIF) SPICE toolkit \citep{Acton1996, Acton2018}. These parameters generally depend on the precise determination of Saturn’s pole, the ring plane radius being the most uncertain parameter with errors of $< 10 \, \mathrm{km}$ for every occultation. Errors in $B$ and $\phi$ are generally less than $1^{\circ}$. Knowing the position and velocity of Mimas enables us to compute the phase of the wave using equation 47a in SCL:

\begin{equation}
  \Phi_{\mathrm{theory}} = (4 \Omega_M + \mu_M)t - 4\Theta - \frac{\pi}{4}
  \label{phase}
\end{equation}

\noindent where $\Omega_M$ and $\mu_M$ are the Keplerian and vertical frequencies of Mimas and $\Theta$ is the azimuthal position of the occultation ($t=0$ at the ascending node of Mimas).  The gravitational moments used to determine the frequencies were taken from \citet{j2}, while the other physical parameters of Saturn and Mimas (mass, semi-major axis of Mimas, and inclination of Mimas) were taken from Jet Propulsion Laboratory Horizons database \citep{jacobson}; these are presented in Table \ref{mimas}. We find a theoretical value for the 5:3 vertical resonance location of $r_V = 131902 \, \mathrm{km}$ and a wave amplitude of $A_V = \frac{472\mathrm{m}}{\sqrt{\sigma_{100}}}$ where $\sigma_{100}$ is the surface mass density in $100 \, \mathrm{g/cm^2}$.

An alternative way of computing the phase is through observations. From the position of the optical depth peaks and troughs in the data a phase can be determined by fixing \citet{gresh}'s best-fit parameters of the SCL model---surface density $\sigma$ and viscosity $\nu$---while letting the phase vary and applying a $\chi^2$ minimization scheme. This data-driven value for the phase is reported in Table \ref{table2} as $\Phi_{\mathrm{obs}}$.

\begin{table}[htbp]
  \fontsize{7pt}{9.25pt}\selectfont
  \addtolength{\tabcolsep}{-1pt}
  \centering
    \begin{tabular}{l|ccccccccccc}
       Occultation & $B \, \textrm{[deg]}$ & $\phi \, \textrm{[deg]}$ & $B_{\mathrm{eff}} \, \textrm{[deg]}$ & $\textrm{Lon. \, [deg]}$ & $Z$ & $\textrm{sign}(\dot Z)$ & $\Phi_{\mathrm{theory}} \textrm{[deg]}$ &$\Phi_{\mathrm{obs}} \textrm{[deg]}$ &$b$ & $I_0$ & $\tau_{\mathrm{max}}$ \\ \hline
      \multicolumn{11}{l}{$B > 60^{\circ}$ and $|\cos{\phi}| > \frac{1}{2} \hspace{5cm}$Complete subset size = 10 occultations} \\  \hline 
    AlpCru(100)I & 68.2  & 154.5 & 70.1  & 213.6 & -5016.5 & 1       &100  &-  &0.50  & 434.18 & 6.0 \\
    AlpCru(92)I & 68.2  & 169.6 & 68.5  & 114.7 & -4232.1 & 1        &160  &-  &0.50  & 518.56 & 6.3 \\
    EpsCas(104)I & -70   & 187.8 & 70.1  & 152.8 & -2034.7 & -1      & 87  &- &0.04  & 4.59  & 3.8 \\
    BetCru(98)I & 65.2  & 192.7 & 65.7  & 319   & 3954.6 & -1        &248  &-  &0.23  & 274.05 & 7.1 \\
    BetCen(104)I & 66.7  & 209.3 & 69.4  & 321   & -1126.9 & -1      &303  &-  &0.31  & 357.50 & 6.7 \\
    BetCen(105)I & 66.7  & 212.9 & 70.1  & 83.1  & 5023  & 1         & 62  &- &0.15  & 313.44 & 5.2 \\
    BetCen(78)E & 66.7  & 45.9  & 73.3  & 206.5 & -5087.4 & -1       &295  &- &0.38  & 564.08 & 7.0 \\
    BetCen(77)E & 66.7  & 47.5  & 73.8  & 12.6  & 4961  & 1          &161  &-  &0.34  & 593.24 & 7.4 \\
    BetCen(92)E & 66.7  & 55    & 76.2  & 75.5  & 4143.3 & -1        &318  &-  &0.23  & 452.62 & 6.2 \\
    EpsCas(104)E & -70   & 122.3 & 79    & 123.6 & -4468.4 & 1       &299  &-  &0.04  & 4.56  & 3.6 \\ \hline
      \multicolumn{11}{l}{$B > 60^{\circ}$ and $|\cos{\phi}| < \frac{1}{2}\hspace{5cm}$Complete subset size = 17 occultation} \\  \hline
    BetCen(77)I & 66.7  & 270.4 & 89.8  & 257.3 & -1823.4 & 1        &164  &-  &0.25  & 587.23 & 7.7 \\
    BetCen(85)I & 66.7  & 277.5 & 86.8  & 41.8  & -4526.9 & -1       &210  &-  &0.71  & 1066.32 & 7.6 \\
    BetCen(89)I & 66.7  & 278   & 86.6  & 322.8 & -572.3 & 1         &277  &-  &0.33  & 498.23 & 7.7 \\
    AlpCru(100)E & 68.2  & 93.9  & 88.5  & 188.8 & 807.8 & 1       &288  &-  &0.50  & 428.76 & 5.8 \\
    BetCen(102)I & 66.7  & 249.1 & 81.3  & 0.9   & 4984.3 & -1       & 53  &- &0.33  & 370.17 & 7.3 \\
    BetCen(104)E & 66.7  & 105.5 & 83.5  & 299.7 & -3402.1 & 1       &154  &-  &0.29  & 342.97 & 8.6 \\
    BetCru(262)I & 65.2  & 257.3 & 84.2  & 18    & 5014.2 & -1       &343  &-  &0.19  & 109.53 & 6.4 \\
    BetCru(253)I & 65.2  & 255.6 & 83.4  & 187.3 & -4002.7 & -1      &157  &-  &0.20  & 106.17 & 6.5 \\
    GamCas(100)E & -66.3 & 72.5  & 82.5  & 93.5  & -3408.3 & 1       &259  &-  &0.09  & 54.16 & 6.1 \\
    GamAra(37)I & 61    & 248.7 & 78.6  & 62.1  & 4684.1 & 1       &133  &-  &0.12  & 53.50 & 4.2 \\ \hline
    \multicolumn{11}{l}{$ 60^{\circ}> B > 30^{\circ}$ and $|\cos{\phi}| > \frac{1}{2}\hspace{4.2cm}$Complete subset size = 35 occultations} \\ \hline
    EpsCen(65)I & 59.6  & 226.6 & 68.1  & 343.1 & -1573.5 & -1       &220  &143  &0.26  & 259.43 & 6.6 \\
    ZetCen(60)I & 53.6  & 228.1 & 63.8  & 258.7 & -4185.7 & -1       &235  &119  &0.23  & 214.36 & 6.6 \\
    DelCen(98)I & 55.6  & 211.3 & 59.6  & 54.8  & 402.3 & 1          &100  & 16  &0.09  & 34.47 & 5.8 \\
    EpsLup(36)E & 51    & 44.9  & 60.2  & 135.3 & 2840.1 & -1        & 99  &125  &0.14  & 65.34 & 5.7 \\
    GamLup(32)E & 47.4  & 33.8  & 52.6  & 158.4 & 5097.4 & 1         &310  &333  &0.15  & 145.17 & 5.4 \\
    LamSco(29)E & 41.7  & 148.3 & 46.3  & 222.1 & -4753.7 & -1       & 35  &  9  &0.32  & 567.90 & 4.6 \\
    ZetPup(171)I & 38.6  & 202.1 & 40.8  & 120.2 & 2022.7 & -1       &350  & 23  &0.12  & 49.51 & 4.3 \\
    EtaLup(34)E & 44.5  & 357.1 & 44.5  & 90    & -3167.9 & -1       &353  & 44  &0.17  & 93.96 & 4.2 \\
    TheAra(40)E & 53.9  & 28.6  & 57.3  & 197.2 & -2036.3 & -1     &269  &218  &0.09  & 24.98 & 4.1 \\
    KapCen(42)I & 48.5  & 168.5 & 49.1  & 237.3 & 228.4 & 1        & 88  &62   &0.24  & 82.26 & 4.1 \\ \hline
    \multicolumn{11}{l}{$60^{\circ}> B > 30^{\circ}$ and $|\cos{\phi}| < \frac{1}{2}\hspace{4.2cm}$Complete subset size = 37 occultations} \\ \hline
    ZetCen(112)I & 53.6  & 239.7 & 69.6  & 112.5 & 4393.8 & -1       &345  &216  &0.03  & 37.31 & 6.4 \\
    AlpAra(32)I & 54.4  & 277.9 & 84.4  & 258.6 & 3837.8 & -1        &311  &311  &0.15  & 76.61 & 6.1 \\
    KapCen(35)E & 48.5  & 85.5  & 86.1  & 302.8 & -5074.8 & 1        &279  &305  &0.12  & 92.03 & 5.8 \\
    TheCar(190)I & -43.3 & 252.4 & 72.2  & 219.3 & -4463.7 & 1     &276  &276  &0.16  & 25.33 & 5.7 \\
    ZetCen(62)E & 53.6  & 70.1  & 75.9  & 8.9   & -4066 & 1          & 46  & 85  &0.19  & 212.18 & 5.5 \\
    TheAra(41)E & 53.9  & 82.4  & 84.5  & 205.4 & 1720.2 & -1        &193  &193  &0.09  & 23.52 & 5.5 \\
    DelCen(64)E & 55.6  & 110.6 & 76.4  & 194.3 & 4773.2 & 1         &327  &353  &0.09  & 52.91 & 5.0 \\
    KapCen(36)I & 48.5  & 241.2 & 67    & 228.6 & -2480.8 & -1       &329  &200  &0.24  & 88.26 & 5.0 \\
    DelPer(37)I & -54   & 264.5 & 86    & 145.7 & 4558.5 & -1        & 28  & 28  &0.08  & 27.26 & 5.0 \\
      AlpLup(248)E & 53.9  & 111.4 & 75.1  & 73.2  & 4024.3 & -1       &150  &175  &0.22  & 16.97 & 4.8 \\
    \end{tabular}%
    
\end{table}%

\begin{table}[htbp]
  \fontsize{7pt}{9.25pt}\selectfont
  \addtolength{\tabcolsep}{-1pt}
  \centering
    \begin{tabular}{l|ccccccccccc}
       Occultation & $B \, \textrm{[deg]}$ & $\phi \, \textrm{[deg]}$ & $B_{\mathrm{eff}} \, \textrm{[deg]}$ & $\textrm{Lon.} \, \textrm{[deg]}$ & $Z$ & $\textrm{sign}(\dot Z)$ & $\Phi_{\mathrm{theory}} \textrm{[deg]}$ &$\Phi_{\mathrm{obs}} \textrm{[deg]}$ &  $b$ & $I_0$ & $\tau_{\mathrm{max}}$ \\ \hline
     \multicolumn{11}{l}{$B < 30^{\circ}$ and $|\cos{\phi}| > \frac{1}{2}\hspace{5cm}$Complete subset size = 11 occultations}  \\ \hline
    GamPeg(172)I & -20.3 & 36.7  & 24.7  & 283.4 & 4683.3  & 1     &328  &339 &0.01  & 11.49   & 1.7 \\
    AlpCMa(274)E*   & 13.5  & 40.4  & 17.5  & 188.9 & 2833.2  & -1    &245  &232 &0.07  & 23.75   & 1.7 \\
    SigSgr(114)I   & 29.1  & 330.2 & 32.6  & 262.3 & -492.7  & 1     &160  &133 &0.04  & 33.50   & 2.7 \\
    AlpVir(34)E    & 17.3  & 332.9 & 19.2  & 259.5 & 2584    & 1     &207  &220 &0.50  & 979.08  & 1.7 \\
    GamPeg(32)I    & -20.3 & 138.6 & 26.2  & 226.6 & 2857.3  & 1     &342  &342 &0.53  & 149.51  & 1.8 \\
    AlpVir(8)I   & 17.3  & 141.1 & 21.8  & 162.3 & 4488.6  & -1    &144  &221 &5.00  & 1000.46 & 1.8 \\
    KapCMa(168)I & 29.3  & 175.9 & 29.4  & 268.8 & -3886.5 & 1     &270  &321 &0.04  & 6.06    & 1.7 \\
    BetCMa(211)I\#  & 14.2  & 223.4 & 19.2  & 67.1  & -3637.3 & 1     &  1  & 232 &0.05  & 43.52   & 1.8 \\
    EpsCMa(276)E   & 26    & 50.6  & 37.5  & 217.7 & 4192    & -1    &108  &173 &0.12  & 69.53   & 2.6 \\
      AlpVir(210)I\#  & 17.3  & 311.5 & 25.1  & 46    & -4015.8 & 1     &259  &104 &0.04  & 138.22  & 2.5 \\ \hline
      \multicolumn{11}{l}{$B < 30^{\circ}$ and $|\cos{\phi}| < \frac{1}{2}\hspace{5cm}$Complete subset size = 20 occultations} \\ \hline
    DelSco(236)I & 28.7  & 263.2 & 77.9  & 10.8  & -5022.7 & -1    &171  &168 &0.65  & 22.98   & 2.8 \\
    AlpVir(232)E & 17.3  & 89.3  & 87.8  & 316.7 & -2068.6 & 1       &284  &284 &0.46  & 125.75  & 2.3 \\
    AlpVir(34)I & 17.3  & 232.8 & 27.2  & 24    & 579.6 & 1          &225  &186 &0.29  & 1021.26 & 2.2 \\
    SigSgr(244)I & 29.1  & 269.3 & 88.8  & 86.6  & -1341.1 & 1       &313  &313 &0.04  & 16.60   & 2.1 \\
    GamPeg(36)E & -20.3 & 66.8  & 43.2  & 343.5 & 3206.1 & 1       & 59  &34 &0.45  & 141.95  & 1.8 \\
    GamPeg(211)E & -20.3 & 122.5 & 34.6  & 207.2 & 518.8 & -1        &200  &200 &0.02  & 13.19   & 1.8 \\
    AlpVir(211)I & 17.3  & 267.2 & 81.2  & 103.3 & -2632.7 & 1       &230  &269 &0.47  & 132.50  & 1.8 \\
    AlpVir(8)E & 17.3  & 91.3  & 85.9  & 197.2 & 3679.9 & -1         & 19  &45 &5.00  & 1053.60 & 1.7 \\
    KapCMa(168)E & 29.3  & 127.3 & 42.8  & 286.8 & -1613.5 & 1       &229  &284 &0.04  & 6.19    & 1.7 \\
    AlpCMa(281)I* & 13.5  & 230.1 & 20.5  & 144.3 & 3652.3 & 1        &143  &195 &0.06  & 27.91   & 1.7 \\
    \end{tabular}%
    \caption{Dataset for this work. We divided the UVIS data set into 6 subsets comprising different geometries, each containing 10 occultations. The theoretical phases of the waves were computed using Eqn. (\ref{phase}) and the latest SPICE kernel for the longitude of the occultation relative to Mimas (the `Lon' column). The observed phases were computed by fitting the phase of the wave in each occultation individually with fixed global parameters $(\beta,\nu,\sigma)$. $Z$ and $\mathrm{\dot{Z}}$ are the vertical position of Mimas with respect to the equator and the sign of its velocity respectively, both are required to compute the phase of the wave. ``$\#$'' indicates that the difference between the theoretical and predicted phase is more than $90^{\circ}$. ``$*$'' indicates that the dispersion relation, or the predicted distances between peaks of optical depth, are inconsistent with the data for these occultations.}%
    \label{table2}%
    
  \end{table}%

\begin{table}
  \centering
\begin{tabular}{ |c|| c c c| } 
 \hline

           & $GM \, [\mathrm{\frac{km^3}{s^2}}]$ & $a \, [\mathrm{km}]$ &$i \, [\mathrm{deg}]$     \\
  \hline
  \hline

  Mimas      &   $2.50349$                      & $185539$ & $1.574$  \\
  
  Saturn      &  $37931206.2$  &  -        & -          \\
  
  \hline

\end{tabular}
\caption{Physical parameters of Mimas and Saturn used in this work. Taken from the JPL Horizons database (accessed in November 2019).}
\label{mimas}
\end{table}

\pagebreak

\section{Ray-tracing model}
\label{sec:4}

The relationship between the slope of the wave and the extinction of starlight in a UV occultation is non-linear, both in the SCL model and in our model. In SCL the shape of the wave enters into the optical depth by determining the limits of the optical depth path integral.

\begin{equation}
  \tau = \int^{l_2}_{l_1} n \bar{\sigma}_{\mathrm{ring}} dl
  \label{tau0}
\end{equation}

\noindent where $n$ is the number density along the light ray, $\bar{\sigma}_{\mathrm{ring}}$ is the average cross-section of absorbers over the path of the light ray, and $l_1$ and $l_2$ are the ring's entry and exit points along the light ray.

Since we use SCL theory to trace the wave profile, $\bar{\sigma}_{\mathrm{ring}}$, $n$, and the thickness of the ring $d$ are uniform throughout the wave \citep{SCL, gresh}. The predicted haze is formed by relatively small particles (see Table \ref{uvisvims}) and hence keeping $n$ uniform is compatible with the haze. A quick calculation shows that, if the haze particles are an order of magnitude smaller than the ring's, they constitute $1\%$ of the mass at the peak of the haze's optical depth. This allows us to write the product $n \bar{\sigma}_{\mathrm{ring}}$ in terms of the normal optical depth outside of the wave $\tau_{n0}$. Consider an occultation along the wave where the light comes only from the direction normal to the plane, then

\begin{equation}
  \tau_{n0} = \int^d_0 n \bar{\sigma}_{\mathrm{ring}} dz = n \bar{\sigma}_{\mathrm{ring}} d
  \label{taun}
\end{equation}

\begin{equation}
  n \bar{\sigma}_{\mathrm{ring}} = \frac{\tau_{n0}}{d}
\end{equation}

  So we can write

  \begin{equation}
   \tau = \int^{l_2}_{l_1} \frac{\tau_{n0}}{d} dl
  \end{equation}

  In order to better match the Voyager I radio occultation, \citet{gresh} introduces a Gaussian enhancement to the optical depth without a physical motivation for producing such an enhancement. We showed in \S \ref{sec:2} that a haze (Eqn. \ref{eureka}) is generated as a consequence of the rotation of self-gravity wakes within the BW; adding this haze to Eqn. (\ref{tau0}) we get:
  
\begin{equation}    
  \tau = \int^{l_2}_{l_1} \frac{\tau_{n0}}{d}  dl + \int^{l_4}_{l_3} \frac{\beta}{D} \left|\frac{dz}{dx}\right|_{\mathrm{max}} dl
  \label{tau}
\end{equation}

\noindent where $l_3$ and $l_4$ are the entry and exit points of the haze along the light ray, $\beta$ is related to the number of particles released by the wakes (see \S \ref{sec:2}) and $D$ is the vertical thickness of the haze. $\left|\frac{dz}{dx}\right|_{\mathrm{max}}$ is the maximum slope over a period at a given radial coordinate $x$. To simplify Eqn. (\ref{tau}) we make use of the azimuthal and north-south symmetry of the rings which makes the integrals depend only on the radial coordinate $x$. Then, if $B$ is the angle between the rings and the light ray and $\phi$ is the angle between the radial direction and the light ray (projected into the ring's mean plane), we can write:

\begin{equation}
  dl = \frac{dx}{\cos{B}\cos{\phi}}
  \label{radial}
\end{equation}

We know the cross-section $\bar{\sigma}_{\mathrm{ring}}$ and the column density $n_{\mathrm{ring}}$ are not isotropic however (they depend on the angle $\phi$), due to self-gravity wakes having a consistent average orientation \citep{Colwell2006}.  Thus, the normal optical depth obtained by simply multiplying the oblique optical depth by $\sin{B}$ is not the true normal optical depth defined in Eqn. (\ref{taun}). Nevertheless, it isn't the goal of our model to describe the dependence of $\bar{\sigma}_{\mathrm{ring}}$ or $n_{\mathrm{ring}}$ on the geometry of the occultation, but only to explain the variation of the optical depth in the wave with respect to the optical depth outside the wave region. Thus, we define the background slanted optical depth as:

  \begin{equation}
    \tau_{0} = \frac{\tau_{n0}}{\sin{B}} = \frac{n \bar{\sigma}_{\mathrm{ring}}d}{\sin{B}}
    \label{slanted}
  \end{equation}

  \noindent where $\tau_{0}$ is determined by averaging values for the optical depth between $r=131600 \, \mathrm{km}$ and $r=131700 \, \mathrm{km}$, a region just outside of the wave.

  By substituting Eqns. (\ref{radial}) and (\ref{slanted}) into Eqn. (\ref{tau}) we get the equation for the predicted optical depth:

\begin{equation}
  \tau = \int^{r_2}_{r_1} \frac{\tau_0}{d} \sin{B} \frac{dx}{\cos{B}\cos{\phi}} + \int^{r_3}_{r_4} \frac{\beta}{D} \left|\frac{dz}{dx}\right|_{\mathrm{max}} \frac{dx}{\cos{B}\cos{\phi}}
  \label{tauf}
\end{equation}

\noindent where the limits of integration $r_1, r_2$ and $r_3, r_4$ are the radial coordinates of the entry and exit points of the light ray as it crosses the ring and the haze respectively. These radial coordinates are computed by numerically finding the intersection of the trajectories of incoming photons, \[z_{\mathrm{photon}}=\sin{B}\frac{x_{\mathrm{photon}}-x_0}{\cos{B}\cos{\phi}},\] with the curves $h(x) \pm \frac{d}{2}$ (for $r_1$ and $r_2$) and the curves delimiting the upper and lower boundary of the haze depicted by the blue fill in Figure \ref{haze} (for $r_3$ and $r_4$). Here $h(x)$ is the wave profile derived in SCL (see Eqn. \ref{h}). To visualize the ray-tracing, consider the light ray (dashed gray line) shown in Figure \ref{haze}. The haze's shape cannot be written in closed form; therefore, in order to trace its upper and lower boundaries we run the numerical procedure described in \S \ref{sec:23} for each occultation.

Contrasting with \citet{gresh}'s second integral in Eqn. (\ref{tauf}), their Gaussian enhancement adds three free parameters that vary independently: width, amplitude, and position. On the other hand, Eqn. \ref{tauf} adds one free-parameter for the haze, $\beta=\Sigma \sigma_{\mathrm{haze}}$, which is related to the number of particles released by the wake as they rotate and collide with neighboring particles (see \S\ref{sec:23}). All the spatial information of the haze---shape, width, and position---is fixed by the wave profile $h$ (Eqn. \ref{h}) which is also heavily constrained by the first integral in Eqn. (\ref{tauf}). In \citet{gresh} the combinations of angles $\frac{\tan{B}}{\cos{\phi}}$ lead to the definition of the angle $\tan{B_{\mathrm{eff}}} = \frac{\tan{B}}{\cos{\phi}}$. Because this combination does not occur in the second term in Eqn. (\ref{tauf}), we prefer to leave the angle $\phi$ explicit.

The shape of the wave itself is a function of the surface mass density $\sigma$, the viscosity $\nu$, and the thickness $d$ which are free parameters. New to our model is free-parameter $\beta$. The quantity to be compared to the data is the transparency, which is the normalized stellar counting rate after background subtraction:

    \begin{equation}
      T_{\mathrm{model}} =  e^{-\tau} = \exp{(-\int^{r_2}_{r_1} \frac{\tau_0}{d} \sin{B} \frac {dx}{\cos{B}\cos{\phi}} - \int^{r_3}_{r_4} \frac{\beta}{D} \left|{\frac{dz}{dx}}\right|_{\mathrm{max}} \frac{dx}{\cos{B}\cos{\phi}})}
      \label{model}
    \end{equation}

    Note that while the thickness $d$ appears directly in the above expression it does not figure as a relevant predictor in our model. As \citet{gresh} discovered, the effects of increasing $d$ on these ray-tracing simulations of bending waves are equivalent to increasing the resolution of the simulation. The optical depth profile becomes smoother in both cases because there are more data points probing the slopes of the ring. Any difference between a thickness of $10 \ \mathrm{m}$ to $100 \ \mathrm{m}$ in the rings would not make a difference in a dataset that is fixed at a resolution of $400 \, \mathrm{m}$. Since the BW occultations at these resolutions can't be used to determine the thickness of the rings directly, we can fix the value of $d$ for all runs, which we choose to be $15 \, \mathrm{m}$. Similarly, the thickness of the haze, which determines the limits of the second integral in Eqn. (\ref{model}), is derived from the dynamics in \S \ref{sec:2} and also has values under the resolution of our data; by the same argument we choose to fix $\epsilon$ (the coefficient of restitution which determines the thickness of the haze) at two values, $0.5$ and $1$. We are then left with a total of three free parameters: the viscosity $\nu$, the surface mass density $\sigma$, and the particles released by the wakes per area times their cross-section $\beta$.

   Note that by taking $\beta$ to be a constant for all occultations we assume that the product $n_{\mathrm{haze}}\sigma_{\mathrm{haze}}$ does not depend on the azimuthal viewing angle $\phi$. This becomes important in the next section as we compare Eqn. (\ref{model}) against the data.

\section{Comparison with data}
\label{sec:5}

The ray-tracing code is used to compute the path integrals in Eqn. (\ref{model}) for $500$ light rays which are drawn at a radial separation of $400 \, \mathrm{m}$ each, covering a radial distance of $200 \, \mathrm{km}$ in the wave region, hence matching the resolution of the data set (the exception to this is EpsCas104E that has one null optical depth value within the BW: the corresponding light ray is removed when simulating this occultation). We use a reduced $\chi^2$ minimization method where we find the best parameter values by comparing the model with the $60$ occultations shown in Table \ref{table2}. The reduced  $\chi^2$ is given by:
    
\begin{equation}
  \chi^2_R = \frac{1}{D_f}\Sigma{\left(\frac{T - T_{\mathrm{model}}}{\Delta}\right)^2}
\end{equation}

\noindent where $\Delta = \frac{\sqrt{I - b}}{I_o}$ and $D_f$ is the degrees of freedom. The uncertainty of $I_0$ and $b$ is negligible due to the small value of $b$ and the amount of data points used to compute them both. $T_{\mathrm{model}}$ is given by Eqns. (\ref{model}), and $T$ is given by Eqn. (\ref{obs}). The wave profile predicted by SCL theory has been modified to include a haze of particles whose shape is computed using the particles' velocity (Eqn. \ref{kick}) and number density (Eqn. \ref{eureka}) with $\beta$ as a free parameter. The total free parameters of our model are $\beta$, the surface density $\sigma$, and the viscosity $\nu$, while the thickness of the ring $d$ and the coefficient of restitution $\epsilon$ have been fixed at values $d = 15 \, \mathrm{m}$ and $\epsilon = 0.5$ and $1$.

\begin{figure}[t]
  \centering
    \includegraphics[width = 0.9\linewidth]{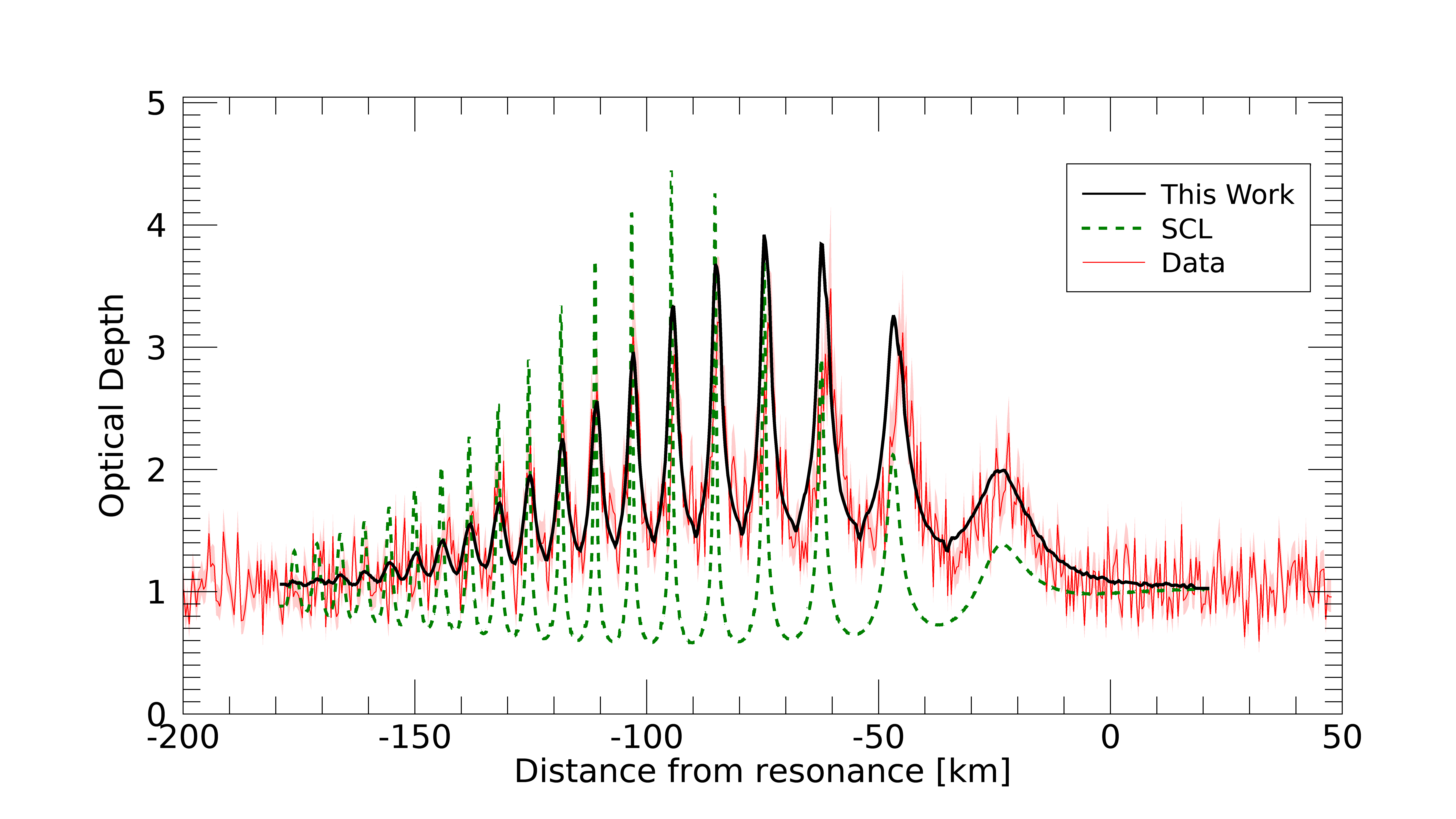}
  \caption{We compare an occultation of $\gamma$-Pegasus (rev 32I, $B = 20.3$,$B_{\mathrm{eff}}= 26^{\circ}$ ) seen by Cassini (solid red) to our model (solid black) with the parameters from the 3rd row in Table \ref{t1}, and SCL theory (dashed green) using the viscosity from \citet{ESPOSITO1} ($\nu = 280 \, \mathrm{cm^2/s}$) and our best-fitted $\sigma$ (our $\sigma$ is used to ensure the peaks between both models better match the data). The error bars in the data are displayed as a light red filled curve in the background of the plot. The underprediction of the troughs in SCL is addressed in the new model, as the haze layer increases the minimum optical depth of the occultation, and the peaks fit better due to the shorter damping length. Note that the predicted peaks are now broader in the new model due to the extended height of the haze, which is something we also see in the data. The discrepancies in the separation of the initial peaks of the model with respect to the data persist.}
  \label{new}
\end{figure}

\begin{figure}[t]
  \centering
  \includegraphics[width= 1\linewidth]{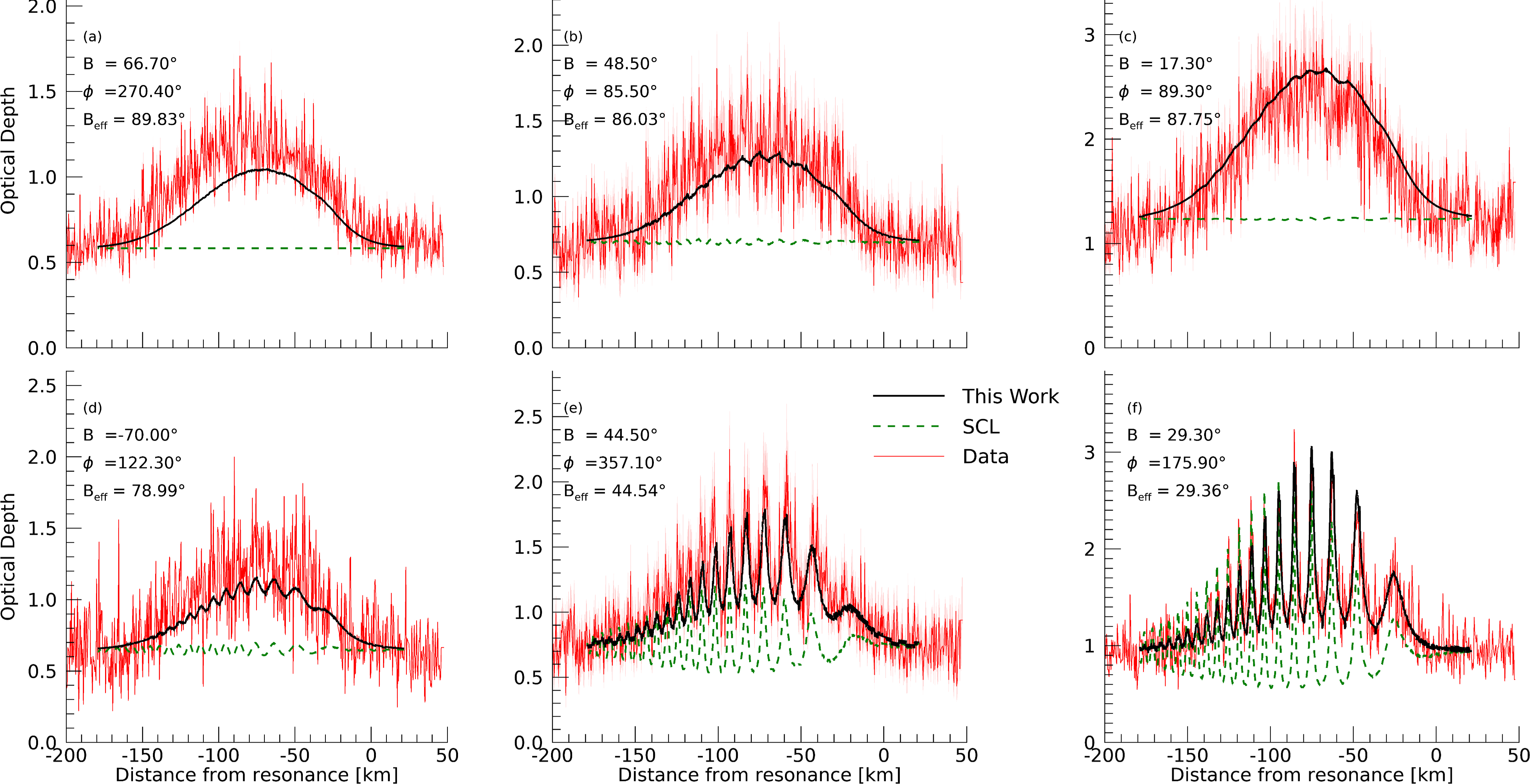}
  \caption{The haze model (3rd row of Table \ref{t1}, solid black lines), and the SCL model (dashed green; same parameters as in Figure \ref{new}) compared to occultations (red lines) of (a) $\beta$-Centauri (rev77I), (b) $\kappa$-Centauri (rev35E) (c) $\alpha$-Virginis (rev232E), (d) $\epsilon$-Cassiopeiae (rev104I) (e) $\eta$-Lupus (rev34E), and (f) $\kappa$-Canis Majoris (rev168I). The error bars are displayed as a light red filled curve in the background of the plot. In the top row we have azimuthal occultations ($\cos{\phi} < 1$), on the bottom row radial occultations ($\cos{\phi}>1$), at high, mid and low $B$ angles (from left to right). For high $B_{\mathrm{eff}}$ occultations, the SCL theory predicts an unchanging optical depth throughout the wave, but instead the data shows a symmetric rise in the optical depth centered at $-80 \, \mathrm{km}$ from resonance. Our model reproduces both the position of the peak and the width of the Gaussian-like enhancement. The amplitude of these peaks, however, does not vary with opening angle $B$ as our model predicts: for azimuthal occultations the scaling with $B$ of the optical depth is too rapid---the model underpredicts (a) and slightly overpredicts (c) the peak of the enhancement---while for radial occultations the scaling works well. The haze model also present a significant improvement when matching the data of low $B_{\mathrm{eff}}$ occultations.}
  \label{fig:33}
\end{figure}

The best-fit parameters for the haze model and the SCL model (no haze), are in Table \ref{t1}. Figures \ref{new} and \ref{fig:33} plot the best-fit optical depth profile for the haze (black line) and SCL (dashed green line) models, and the observed UVIS occultations (red line) for 7 occultations.

Table \ref{t1} shows that the models with the haze ($\beta \ne 0$) fit significantly better than the SCL model for the dataset. The improvement for low $B_{\mathrm{eff}}$ is represented in Figure \ref{new}. Most notably, Figure \ref{new} shows that the troughs of the optical depth pattern now coincide with the predicted optical depth due to the Gaussian-like enhancement produced by the haze. Moreover, the bases of the peaks are broader than those of SCL theory and resemble the observed shape of the optical depth profile. The broadening of the peaks' bases is caused by the thickening of the haze at the minima and maxima of the wave (shown in \S \ref{sec:23}). Note, however, that the position of the peaks are unchanged by our modification, as we still use SCL's dispersion relation. Figure \ref{new} also shows that the damping length due to the viscosity of $\nu = 576 \, \mathrm{cm^2/s}$ closely matches the observed length of the wave. This high viscosity increases the predicted value for the slope of the wave which is $17^{\circ}$ for our best-fit model.

  Table \ref{t1} shows that using the observed phase instead of the theoretical phase improves the fit, but it does not change our values for $\beta$, $\nu$, and  $\sigma$ significantly. Likewise, changing the coefficient of restitution, which changes the thickness of the haze (see Figure \ref{haze}), does not alter the agreement with the data.

\begin{table}[b]
  \centering
\begin{tabular}{ |c c c|| c| c| c| c|} 
 \hline
        Model &Phase& $\epsilon \, \mathrm{[-]} $ &$\chi^2_R \, \mathrm{[-]}$ & $\sigma \, \mathrm{[\frac{g}{cm^2}]} $ & $\nu \, \mathrm{[\frac{cm^2}{s}]}$ & $\beta \, [-]$  \\
  \hline
  \hline
        SCL  &$\Phi_{\mathrm{theory}}$ & $-$& $22.15$    & $36.4 \pm 5$& $3540 \pm 40$     & $-$    \\
  \hline
  Haze&$\Phi_{\mathrm{theory}}$& $1$& $5.56$ & $37.0 \pm 0.4 $  & $ \ 623 \pm 3 $            & $1.48$  \\ \hline

  Haze&$\Phi_{\mathrm{obs}}$ &$1$&$4.46$ & $36.7 \pm 0.3 $      & $\, \, 576 \pm 3$   & $1.39$  \\ \hline

  Haze&$\Phi_{\mathrm{obs}}$ &$0.5$&$4.48$ & $36.6 \pm 0.3 $      & $\, \, 576 \pm 3 $   & $1.37$  \\ \hline  

\end{tabular}
\caption{Best fit results for the SCL and haze models. There is a significant improvement in the fit of the haze model over the SCL model. Error bars show the $1\sigma$ of the likelihood given by $e^{-\chi^2/2}$. The haze model has 3 free parameters, so all have $D_f=29819$ degrees of freedom. SCL has 2 free parameters and hence $D_f=29879$.}
\label{t1}
\end{table}

In Figure \ref{fig:33} we present 6 occultations, once for each of the data subsets in Table \ref{table2}. The occultations range from higher to lower $B$ from left to right in the figure. Our model represents a significant improvement over SCL for both radial (lower row) and azimuthal (upper tow) occultations. For higher $B_{\mathrm{eff}}$ occultations, the SCL theory predicts little change in the optical depth, while the haze model produces the shape of the optical depth enhancement seen in the data. Nevertheless, we see in the top row that the scaling with $B$ for the haze optical depth, which is simply $1/\sin{B}$ in our model, fails to reproduce the amplitude of this optical depth enhancement for some azimuthal occultations. While the amplitude fits well for intermediate $B$ (b), for high and low $B$ angles our haze model underpredicts (a) and overpredicts (c) the maximum of the smoothed optical depth by more than $10\%$. This effect is not evident in the radial occultations, as the troughs of the optical depth pattern match the data in (e) and (f), while (d) is only slightly below the data's amplitude value. This is a clear example of a $\phi$-dependent optical depth (i.e. azimuthal brightness asymmetry). In other words, the haze does not behave as an isotropic absorbing medium. A potential explanation for this is discussed in the next section.

\section{Discussion}      \label{sec:6}

Discrepancies between the optical depth predicted by the SCL model and those measured by Cassini UVIS and Voyager I \citep{gresh}, have been partially addressed by our modifications to the theory (Figure \ref{new}). The addition of haze particles released by self-gravity wakes changes the wave profile, addressing problem \textit{(1)} of the introduction (\S \ref{sec1}), and improving the likelihood considerably (Table \ref{t1}). Most of the explanatory power in the model comes from the existence of the extra layer of particles.

However, Figure \ref{fig:33} shows that the observed haze optical depth presents azimuthal brightness asymmetry ($\phi$-dependence). Given that the haze is hypothesized to come from self-gravity wakes (which are responsible for the $\phi$-dependence of the flat ring’s optical depth), the observed anisotropy may be inherited from them. To improve our model to explain this anisotropy, a more-in-depth analysis of the dynamics presented in \S \ref{sec:2} is called for. If the self-gravity wakes have a preferred direction inside the BW, and the haze comes from said wakes, the haze will present an anisotropy; which, in turn, can be compared against the observations as an additional test for the dynamical model. Preliminarily, Figure \ref{wakepotential} suggests that the azimuthal direction may be a preferred orientation, but further analysis is needed to determine the attractors in the rotation of wakes in the BW.

\begin{figure}[t]
  \centering
  \includegraphics[width = 0.6\linewidth]{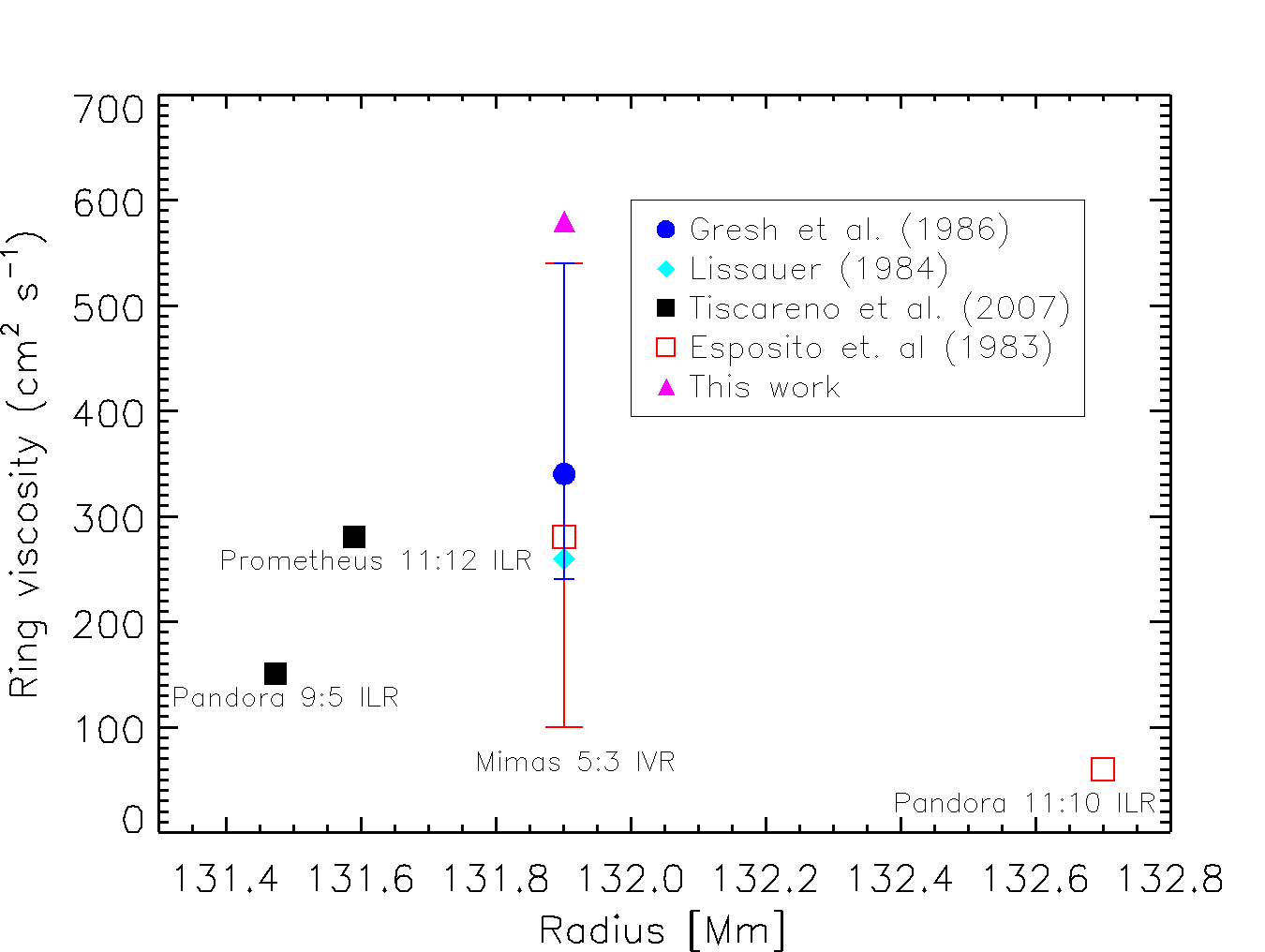}
  \caption{Measurements of the viscosity of the rings made by various authors around the bending wave region. The measurements done using the Mimas BW are the ones at $131.9 \, \mathrm{Mm}$. The values computed in this paper are the filled triangles. This work's values are considerably higher than the more recurrent value of $\nu = 280 \, \mathrm{cm^2/s}$. Our measurement lies outside the error bars of \citet{gresh} and \citet{ESPOSITO1} (\citealt{Tiscareno} and \citealt{Lissauer} didn't include error bars). \citet{ESPOSITO1}, \citet{gresh}, \citet{Tiscareno} and \citet{Lissauer} were using one occultation (or image) for their estimates; in this work we use 60 occultations at different geometries.}
  \label{visc}
\end{figure}

Figure \ref{new} shows that the haze model's fit of the observed wave damping length yields a relatively high kinematic viscosity, partially addressing problem \textit{(3)} of the introduction (\S 1). The best-fit viscosity value more than doubles the viscosity obtained from neighboring density waves \citep{Tiscareno}, and it falls outside the upper error bars in \citet{gresh} and \citet{ESPOSITO1}. We compare our value with the literature in Figure \ref{visc}

Figure \ref{visc} shows that \citet{gresh}, \citet{Lissauer}, and \citet{ESPOSITO1} measured a lower viscosity than the best-fitted values of our work. \citet{ESPOSITO1} used one occultation of $\delta$-Scorpii, while \citet{gresh} used one low-angle radio occultations of the Earth (at two frequencies). \citet{Lissauer} used images of the shadows cast by the elevated ring to estimate where the wave-train ended. In this work we use 60 occultations at better resolution (400 m vs. 2 km). The signal-to-noise is also higher on average since we are looking at brighter sources, like $\gamma$-Pegasus and $\beta$-Centauri.

We see two reasons why their viscosity is lower. The first one pertains to the noise and the spatial resolution of the data: they didn't fit all the cycles because they couldn't see them, so instead of fitting the complete wave-train to find the viscosity they attempted to fit the damping length $\xi_D$, which is the point at which the amplitude of the wave drops a factor of $e$. This approach would work if the SCL theory were adequate for this BW, which we see it isn't in its entirety, as shown in the first row of Table \ref{t1} and Figures \ref{new} and \ref{fig:33}. This leads us to our second reason, the lack of a proper model for the extra optical depth. Without the haze our value for the viscosity would underpredict all of the peaks in \ref{new}. Moreover, in the case of \citet{gresh}, the occultations were so low angle that the optical depth of the peak was taken to be infinity (the slope of the wave and the viewing angle were almost the same), which means they mostly fit the troughs and not the peaks and we can see in Figure \ref{new} that the underdamping in the SCL model's fit is more evident in the peaks than in the troughs. In summary, \citet{gresh}, \citet{Lissauer} and \citet{ESPOSITO1} viscosity values are lower than the ones suggested in this work because in our occultations the late part of the wave-train can be clearly distinguished from the noise, and their prediction for the damping length relies on the SCL theory.

     Nevertheless, the value \citet{gresh} computed for the region is still higher than what \citet{Tiscareno} calculated from analyzing the 11:12 Prometheus density wave which used Cassini data and looks at weak bending waves where SCL is adequate (although \citealt{Tiscareno} points out that the Prometheus 11:12 shows some non-linearity, it isn't enough to expect his results to be inaccurate, given their consistency with other estimates of the viscosity in the region). To be consistent with \citet{Tiscareno} we then have to consider that the BW itself increases the viscosity in a way that a density wave does not.

     To further address problem \textit{(3)}, we need to account for the back-reaction of the wave to the wake's rotation since we can expect that the inclusion of the wake-wave interaction will affect the damping length of the theory. A way to include this added dissipation is to consider the effects that the collisional forces (Eqn. \ref{force1}) have on the wakes' vertical motion. In our current model the collisions are taken to be north-south symmetric in the frame of the wake: the same amount of particles are hitting the wake from above and from below, which results in a net vertical force of zero. This is probably not true, and the inelastic wake-wave collisions would cause a net damping of the vertical motion of the wakes, and of the ring particles in general. Since we haven't included this effect, our model accounts for the extra dissipation by having a high value in the kinematic viscosity of the SCL term, in which the viscous effects of vertical collisions are modeled simply by $F_{\mathrm{visc}}/m = \nu d\dot{z}/dx$ (where $m$ is the mass of the affected particle). However, even our simple collision model suggests a more complex interaction that depends on the slope, on the mass colliding with the self-gravity wake, and on the wake's size. The computation of this viscous force and the subsequent modification to SCL's wave profile are left to future work.
     
     Angular momentum conservation between the rings and satellites indicates that the viscosity responsible for the spreading of the rings, and that which damps the waves, may not be related in a simple fashion \citep{Rawadan}. However, if the momentum transport in both cases is dominated by the motions of self-gravity wakes a higher damping viscosity will be closely related to a more efficient momentum transport in that region. Then, while the Mimas 5:3 BW is relatively strong (the 2nd strongest BW in the rings), self-gravity wakes allow the rings to transport the inflow of energy at resonance and form a continuous wave. Despite this extra dissipation, the Mimas 5:3 BW still has a slope of $\sim 15^{\circ}$, hence we can speculate that without the enhancement in viscosity provided by the self-gravity wakes, the rings would rift and create a gap similar to the ones seen in the Titan 0:-1 nodal BW \citep{nicholson} and in protoplanetary disks misaligned with their star's obliquity \citep{Rowther}. Without a theory of when (at what slope and at what efficiency of viscous transport) warped disks break however, this remains a speculation. An attempt at modeling the Titan 0:-1 nodal BW using Cassini data may shed light on a general theory of rifted warps in disks.
       
       The more efficient angular momentum transport within the wave could also be causing the diffusion of material into the neighboring regions. Consider the values measured by \citet{Tiscareno} of the average surface mass density of the Prometheus 12:11 wave, which lies about $100$ km closer to Saturn than the Mimas 5:3 BW. The average measured surface density value is $\sigma = 46.1 \, \mathrm{g/cm^2}$. This corresponds to an increase of $25\%$ with respect to our best-fit BW value (and a $24\%$ increase with respect to \citealt{gresh}'s best-fit value). The non-linear nature of this density wave could be changing the local value of the surface density \citep{borderies}, but the Prometheus 12:11 is not strong enough for this effect to change it more than $20\%$. So why does $\sigma$ increase in this region? A possible explanation is that density waves, by their very nature of being density-changing phenomena, are unreliable measures for $\sigma$, although this wouldn't explain why the measured values are overstimating the BW value. Another more plausible explanation is that the higher viscosity in the BW region is causing the migration of material. A local radial increase (decrease) in viscosity has been shown to correspond to a local decrease (increase) in surface density in protoplanetary disks \citep{lyra}. If the change in viscosity is too abrupt, it can lead to an instability that can excite the growth of vortices in the disk \citep{pap2}. While this is not the case here, the increase in viscosity in the bending wave region may be causing a decrease in surface density there, and a corresponding increase in the nearby regions.

    Comparing the values of $\Phi_{\mathrm{theory}}$ and $\Phi_{\mathrm{obs}}$ we see a general agreement for the theoretical phase of the wave (Table \ref{table2}). While there are many waves with a phase difference of $\sim 20^{\circ}$, this is to be expected considering that a radial offset of $10 \, \mathrm{km}$, either in the radial vector or the resonance position, and an error of $1^{\circ}$ in the longitude of Mimas translates into a possible phase difference of $25^{\circ}$ between theory and observation. Nevertheless, two stars show an offset in the phase greater than $90^{\circ}$: AlpVir(210)I and BetCMa(276)E. Note that the dispersion relation works well for these occultations once the offset in the phase is corrected. That this only occurs in two out of $40$ occultations makes us hesitant to propose a mechanism driving such a big offset. On the other hand, the offset could be due to a bigger systematic error in the radial offset of these particular occultations.

    While the used SCL dispersion relation does not improve the position of the first peaks of the wave (problem \textit{(2)} of the introduction \S 1) it does work fairly well with most occultations (see Figure \ref{fig:33}). Moreover, the broader shape of the optical depth peaks partially alleviates some of the discrepancies in the first peaks. Nevertheless, we find two low-angle occultations in which the dispersion relation fails not only at the first cycles but everywhere in the wave-train: AlpCMa(274)E and AlpCMa(281)I. A possible cause of the dispersion relation difference can be a radial variation in the surface mass density of the ring. By using the fitted value for the isotropic model $\beta =1.39$, via equation $\beta = \Sigma_0 \bar{\sigma}_{\mathrm{haze}}$ (\S \ref{sec:23}), assuming $\bar{\sigma}_{\mathrm{haze}} = 1 \, \mathrm{cm^2}$ we estimate a maximum haze surface mass density of $0.4 \, \mathrm{g/cm^2}$ which is about $1\%$ the best-fit for the ring's surface mass density. The existence of the haze is therefore consistent with SCL's dispersion relation which assumes a constant surface mass density throughout the wave and, by itself, is not evidence that there is a significant change in the ring's density.

    Nevertheless, if the average cross-section of haze particles is bigger ($10 \, \mathrm{cm}$, although we note this is contrary to what is suggested by Table \ref{uvisvims}) and radially changing (there are bigger particles released when the slope is bigger) there may still be a slow radial change of $5$ to $10\%$ in the surface density which would affect the dispersion relation appreciably in low $B_{\mathrm{eff}}$ angles occultations. Note that incorporating the wake-wave interaction in the bending wave equation might also change the dispersion relation of the wave, potentially improving the predicted wavelength near resonance.

    By gaining a better understanding of the Mimas 5:3 BW, we can help address other problems in planetary rings science. First, consider the unexplained high viscosity of the Mimas 5:3 density wave \citep{borderies}. The source of this viscosity may also be related to the rotation of self-gravity wakes. Local increases in surface density may reduce the local shear-rate at that location, causing wakes to rotate as they adjust their pitch-angle (see Figure \ref{shear}). While the rotation due to a changing shear-rate will be more gentle than the one described in this work, it can contribute to the increase in local viscosity seen in this density wave. Secondly, consider the puzzle set by \citet{Tiscareno2013b} for the Iapetus -1:0 BW. This BW propagates outwards through an abrupt increase in optical depth called `the inner A-ring edge.' Historically, researchers have interpreted this increase in optical depth as an increase in the surface density $\sigma$. However, $\sigma$ also affects the dispersion relation of waves, and no such abrupt change is seen in the wavelength of the Iapetus -1:0 BW. If this edge corresponds to a point where self-gravity wakes to start form, given how unstable self-gravity wakes are to vertical perturbations within BWs, a haze of particles may contribute to the slanted optical depth increase. Properly applying the rigid-bar and haze models to these issues is left for future work.

  \section{Conclusion}      \label{sec:7}

     We have shown that a necessary consequence of introducing rigid self-gravity wakes to the linear bending wave theory is that bending waves generate an extra layer of particles that is proportional to the amplitude of the slope. Even if we consider the scenario of the wake's partial disruption due to the interaction with the bending wave, this disruption stills create the same layer of haze particles, given that the relative vertical velocities between the wakes and the ring are comparable to the characteristic vertical velocities of the particles as the wave propagates.

     Using a ray-tracing code, we tested this additional signal and found that the extra layer of particles has considerable explanatory power for the Cassini UVIS occultations of the BW compared to \citet*{SCL}. Our best model explains the most discrepant and surprising features of the Mimas 5:3 BW; namely, the enhancement of the signal for the cases of normal occultations and a bigger than expected viscosity, $\nu = 576 \, \mathrm{cm^2/s}$, more than two times bigger than the viscosity computed from density waves. This shows that self-gravity wakes can be especially effective in preventing the opening of gaps in a vertically perturbed disk.

     The rigid-bar model for self-gravity wakes proved to be successful not only in the case of the bending wave but also in the flat ring, as it showed that wakes tend to align at the same pitch-angle due to their mutual self-gravity. Thus, this can explain why the average pitch-angle of the wakes is the same throughout the A and B rings even if the ring properties, like surface mass density and distance from Saturn, change considerably within them. The rigid-bar model can be used as an analytical tool to investigate the motion of self-gravity wakes in a perturbed environment, such as density waves.

     While improvements have been made in explaining the Cassini UVIS dataset for the Mimas 5:3 BW, there are still issues with the current bending wave theory and we suggest a path to modify it in a way consistent with the results of this paper. We find that the theoretical dispersion relation tends to fail in the first cycles of the wave, and in some cases, it fails through the entire wavetrain. The reason for this may be the back-reaction that the self-gravity wakes' motion has on the propagation of the wave. The inclusion of this effect is suggested as a path to further improve the theory.

    \section*{Acknowledgements}
     This document is the results of the research project funded by NASA FINESST, award No 80NSSC20K1379.

     \appendix
     \section{Complete form of the torques on self-gravity wakes}
     \label{AA}

  Let $\mu_G = \frac{GM_{S}}{r^3}$ where $r$ is the radial distance from Saturn, $M_{S}n$ is the mass of Saturn and $G$ the gravitational constant.

     \begin{equation}
\begin{array}{l}
\tau_{\mathrm{tidal + wake};x'}=4\big(\overbrace{\mu_G}^{\text{tidal }}-\overbrace{\pi G \sigma k^2 \cos ^2 \theta_w}^{\text{wake }}\big) \frac{L^3}{24} W H \rho_H\left(\hat{z}^{\prime} \cdot \hat{x}\right)\left(\hat{y}^{\prime} \cdot \hat{x}\right) \\
  -2\big(\overbrace{\mu_G}^{\text{tidal }}+\overbrace{2 \pi G \sigma k^2 \sin ^2 \theta_w}^{\text{wake }}\big) \frac{L^3}{24} W H \rho_H\left(\hat{z}^{\prime} \cdot \hat{y}\right)(\hat{y^{\prime}} \cdot \hat{y}) \\
\overbrace{-4 \pi G \sigma k \cos \theta_w \sin \theta_w}^{\text{wake }} \frac{L^3}{24} WH \rho_H\left[(\hat{z^{\prime}} \cdot \hat{y})(\hat{y^{\prime}} \cdot \hat{x})+\left(\hat{z}^{\prime} \cdot \hat{x}\right)\left(\hat{y}^{\prime} \cdot \hat{y}\right)\right]
\end{array}
\end{equation}

\begin{equation}
\begin{array}{l}
\tau_{\mathrm{tidal + wake};y'}  =-4\big(\overbrace{\mu_G}^{\text{tidal }}-\overbrace{\pi G \sigma k^2 \cos ^2 \theta_w}^{\text{wake }}\big) \frac{W^3}{24} L H{\rho_H}\left(\hat{z}^{\prime} \cdot \dot{x}\right)\left(\hat{x}^{\prime} \cdot \hat{x}\right) \\
 +2\big(\overbrace{\mu_G}^{\text{tidal }}+\overbrace{2 \pi G \sigma k^2 \sin ^2 \theta_w}^{\text{wake }}\big) \frac{W^3}{24} L H \rho_H\left(\hat{z}^{\prime} \cdot \dot{y}\right)\left(\hat{x}^{\prime} \cdot \hat{y}\right) \\
 +\overbrace{4 \pi G \sigma k \cos \theta_w \sin \theta_w}^{\text{wake }} \frac{W^3}{24} \rho_H L H\left[\left(\hat{z}^{\prime} \cdot \hat{y}\right)(\hat{x} \cdot \hat{x})+\left(\hat{z}^{\prime} \cdot \hat{x}\right)\left(\hat{x}^{\prime} \cdot y^{\prime}\right)\right]
\end{array}
\end{equation}

\begin{equation}
\begin{array}{l}
\tau_{\mathrm{tidal + wake};z'}=4\big(\overbrace{\mu_G}^{\text{tidal }}-\overbrace{\pi G \sigma \cos^2 \theta_w}^{\text{wake }}\big) \frac{L^3}{24} \rho_{H}HW\left(\hat{x}^{\prime} \cdot \hat{x}\right)\left(\hat{y}^{\prime} \cdot \hat{x}\right) \\
  +2\big(\overbrace{\mu_G}^{\text{tidal }}+ \overbrace{2 \pi G \sigma \sin^2 \theta_w}^{\text{wake }}\big) \frac{L^3}{24} \rho_H H W\left(\hat{x}^{\prime} \cdot \hat{y}\right)\left(\hat{y}^{\prime} \cdot \hat{y}\right) \\
+\overbrace{4 \pi G \sigma k \cos \theta_w \sin \theta_w}^{\text{wake }} \frac{L^3}{24} \rho_HHW\left[(\hat{x} \cdot \hat{y})\left(\hat{y}^{\prime} \cdot \hat{x}\right)+(\hat{x} \cdot \hat{x})\left(\hat{y}^{\prime} \cdot \hat{y}\right)\right]
\end{array}
\end{equation}

Where $W$, $H$ and $L$ are the length of the principal axes of the wake. Centered on the center of the wake, we set a coordinate systems along the principal axes ($x'$, $y'$, $z'$, respectively). $k$ is the wavenumber of the self-gravity wakes, which are treated as a plane-wave field. $\rho_H$ is the density of the self-gravity wake and $\sigma$ is its surface density. $\theta_w$ is the angle with respect to the azimuthal of the long axis $L$ of the oriented wake field.

Consider the direction $\hat{\theta}$, which is the direction of the velocity of a particle colliding using the Hill equations (see Figure \ref{zerovel}). Consider moreover, that the space density of colliding particles is given by $\rho(z) = e^{-(\frac{z}{z_0})^2} $ where $z_0$ is the rings' half-thickness, and note that the $z$ coordinate can be written in terms of $(x',y',z')$. We can then write the torques due to collisions of particles with a space density $\rho_s$, both due to the BW shear (caused by the slope of the bending wave $\frac{dz}{dx}$) and due to the Keplerian shear, as:

\begin{equation}
\begin{aligned}
\tau_{\mathrm{Kep + BWsh};x'}= & 2*(1+\epsilon)\int^{b_{\mathrm{max}}}_{b_{\mathrm{min}}} \biggr|\overbrace{\vec{v}_\theta \cdot \hat{z}'}^{\text{Keplerian }}+(\overbrace{\frac{dz}{dx}(\hat{y}^{\prime} \cdot \hat{x})(\hat{z}^{\prime} \cdot \hat{z})}^{\text{BW shear }}-\omega_{x'}) y^{\prime}(b) \biggr| \\
& +\biggr[\overbrace{\vec{v}_\theta \cdot \hat{z}'}^{\text{Keplerian }}+(\overbrace{\frac{dz}{dx}(\hat{y}^{\prime} \cdot \hat{x})(\hat{z}^{\prime} \cdot \hat{z})}^{\text{BW shear }}+\omega_{y'}) y^{\prime}(b)\biggr] y'(b) \int^{W/2}_{-W/2}\rho_s(z(x',y'(b)))dbdx'
\end{aligned}
\end{equation}

\begin{equation}
\begin{aligned}
\tau_{\mathrm{Kep + BWsh};y'}= & -2*(1+\epsilon)\int^{b_{\mathrm{max}}}_{b_{\mathrm{min}}} \biggr|\overbrace{\vec{v}_\theta \cdot \hat{z}'}^{\text{Keplerian }}+(\overbrace{\frac{dz}{dx}(\hat{x}^{\prime} \cdot \hat{x})(\hat{z}^{\prime} \cdot \hat{z})}^{\text{BW shear }}+\omega_{y'}) x^{\prime}(b) \biggr| \\
& +\biggr[\overbrace{\vec{v}_\theta \cdot \hat{z}'}^{\text{Keplerian }}+(\overbrace{\frac{dz}{dx}(\hat{x}^{\prime} \cdot \hat{x})(\hat{z}^{\prime} \cdot \hat{z})}^{\text{BW shear }}+\omega_{y'}) x^{\prime}(b)\biggr] x'(b) \int^{L/2}_{-L/2}\rho_s(z(x'(b),y'))dbdy'
\end{aligned}
\end{equation}

\begin{equation}
\begin{aligned}
\tau_{\mathrm{Kep + BWsh};z'}= & -2*(1+\epsilon)\int^{b_{\mathrm{max}}}_{b_{\mathrm{min}}} \biggr|\overbrace{\vec{v}_\theta \cdot \hat{x}'}^{\text{Keplerian }}+(\overbrace{\frac{dz}{dx}(\hat{y}^{\prime} \cdot \hat{x})(\hat{z}^{\prime} \cdot \hat{z})}^{\text{BW shear }}+\omega_{z'}) y^{\prime}(b) \biggr| \\
& +\biggr[\overbrace{\vec{v}_\theta \cdot \hat{z}'}^{\text{Keplerian }}+(\overbrace{\frac{dz}{dx}(\hat{y}^{\prime} \cdot \hat{x})(\hat{z}^{\prime} \cdot \hat{z})}^{\text{BW shear}}+\omega_{z'}) y^{\prime}(b)\biggr] y'(b) \int^{H/2}_{-H/2}\rho_s(z(y'(b),z'))dbdz'
\end{aligned}
\end{equation}

Finally the torque caused by the radial gradient of the vertical acceleration due to the bending wave:
\begin{equation}
 \vec{\tau}_{\mathrm{BWg},x'} = -\frac{2}{3}\omega'^2 \frac{\partial z}{\partial y'} (\hat{z} \cdot \hat{z}') (\hat{y}' \times \hat{z}')\rho_H HW\left(\frac{L}{2}\right)^3
\end{equation}

\begin{equation}
  \vec{\tau}_{\mathrm{BWg},y'} = -\frac{2}{3}\omega'^2 \frac{\partial z}{\partial x'} (\hat{z} \cdot \hat{z}') (\hat{x}' \times \hat{z}') \rho_HHL\left(\frac{W}{2}\right)^3
\end{equation}

\begin{equation}
  \vec{\tau}_{\mathrm{BWg},z'} = -\frac{2}{3}\omega'^2 \frac{\partial z}{\partial y'} (\hat{z} \cdot \hat{x}') (\hat{y}' \times \hat{x}') \rho_HHW\left(\frac{L}{2}\right)^3
\end{equation}


     \pagebreak

     \section{Table with the UVIS occultations of the Mimas 5:3 bending wave and their geometry}
     \label{AB}

     The below table collects all UVIS occultations with $\tau_{\mathrm{max}} > 1.5$ for the Mimas 5:3 bending wave region.

     \begin{table}[htbp]
       \fontsize{6pt}{7pt}\selectfont
       \addtolength{\tabcolsep}{0pt}
       \centering
       \begin{tabular}{l|cccccccccc}
         Occultation & $B \, \textrm{[deg]}$ & $\phi \, \textrm{[deg]}$ & $B_{\mathrm{eff}} \, \textrm{[deg]}$ & $\textrm{Lon. \, [deg]}$ & $Z$ & $\textrm{sign}(\dot  Z)$&$b$ & $I_0$ & $\tau_{\mathrm{max}}$ & $\Phi_{\mathrm{theory}} \textrm{[deg]}$ \\ \hline
         AlpAra(105)E & 54.4  & 77.1  & 80.9  & 125.7 & -4046 & -1    & 9950  & 16.94937 & 3.2   & 224 \\
    AlpAra(105)I & 54.4  & 355.8 & 54.5  & 314.9 & 4072  & -1    & 11381 & 18.1537 & 3.5   & 83 \\
    AlpAra(32)I & 54.4  & 277.9 & 84.4  & 258.6 & 3837.8 & -1    & 38303 & 75.4764 & 6.1   & 312 \\
    AlpAra(33)I & 54.4  & 278   & 84.3  & 12.4  & 2512  & 1     & 36362 & 84.57951 & 5.5   & 115 \\
    AlpAra(35)E & 54.4  & 116.1 & 72.6  & 343.7 & -4739.8 & -1    & 36982 & 75.47634 & 3.8   & 268 \\
    AlpAra(63)E & 54.4  & 106.8 & 78.3  & 221.1 & 3969.2 & -1    & 28626 & 38.80406 & 5.5   & 280 \\
    AlpAra(79)I & 54.4  & 6.5   & 54.6  & 207.1 & 688.9 & 1     & 24713 & 32.6135 & 3.6   & 34 \\
    AlpAra(85)E & 54.4  & 93.4  & 87.6  & 85.8  & 5119.7 & 1     & 24444 & 24.13635 & 3.5   & 61 \\
    AlpAra(85)I & 54.4  & 5.7   & 54.6  & 233.9 & 2256.6 & 1     & 24897 & 28.06419 & 3.5   & 306 \\
    AlpAra(86)E & 54.4  & 94.1  & 87    & 147.2 & 2714.7 & 1     & 21499 & 26.68124 & 3.5   & 118 \\
    AlpAra(86)I & 54.4  & 5     & 54.5  & 299.1 & -2625.6 & 1     & 22373 & 28.52165 & 3.5   & 348 \\
    AlpAra(90)E & 54.4  & 93.7  & 87.3  & 103   & 4970.2 & -1    & 20224 & 21.15087 & 3.6   & 7 \\
    AlpAra(90)I & 54.4  & 5.2   & 54.5  & 252.2 & 3600.4 & 1     & 20057 & 27.9169 & 3.5   & 251 \\
    AlpAra(96)E & 54.4  & 82.6  & 84.7  & 124.1 & 3156.1 & -1    & 19349 & 19.06265 & 3.5   & 141 \\
    AlpAra(96)I & 54.4  & 12.8  & 55.1  & 254.8 & 4942.7 & 1     & 19563 & 41.38505 & 3.6   & 271 \\
    AlpAra(98)E & 54.4  & 75.4  & 79.8  & 110.7 & 131.2 & -1    & 17371 & 56.24169 & 3.6   & 231 \\
    AlpAra(98)I & 54.4  & 8.4   & 54.7  & 250.9 & 4825.8 & -1    & 17202 & 65.79578 & 3.6   & 321 \\
    AlpCMa(274)E & 13.5  & 40.4  & 17.5  & 188.9 & 2833.2 & -1    & 23747 & 68.66385 & 1.7   & 246 \\
    AlpCMa(281)I & 13.5  & 230.1 & 20.5  & 144.3 & 3652.3 & 1     & 27911 & 57.7524 & 1.7   & 143 \\
    AlpCru(100)E & 68.2  & 93.9  & 88.5  & 188.8 & 807.8 & 1     & 428757 & 501.3685 & 5.8   & 289 \\
    AlpCru(100)I & 68.2  & 154.5 & 70.1  & 213.6 & -5016.5 & 1     & 434175 & 501.3685 & 6     & 102 \\
    AlpCru(92)I & 68.2  & 169.6 & 68.5  & 114.7 & -4232.1 & 1     & 518555 & 501.3384 & 6.3   & 160 \\
    AlpLup(248)E & 53.9  & 111.4 & 75.1  & 73.2  & 4024.3 & -1    & 16967 & 223.1837 & 4.8   & 150 \\
    AlpLyr(175)I & -35.2 & 233.8 & 50.1  & 318.5 & 1522.8 & -1    & 8139  & 21.61554 & 4     & 104 \\
    AlpLyr(202)E & -35.2 & 35    & 40.7  & 191.8 & -4393 & -1    & 8139     & 21.61554     & 3.7   & 147 \\
    AlpLyr(202)I & -35.2 & 236   & 51.6  & 90.2  & 3092.6 & -1    & 6485  & 19.1991 & 2.9   & 277 \\
    AlpLyr(206)I & -35.2 & 252.9 & 67.4  & 151.7 & -1264.5 & -1    & 6801  & 17.57583 & 3.5   & 82 \\
    AlpScoB(13)E & 32.2  & 117.8 & 53.4  & 255.1 & -831.6 & -1    & 3409  & 104.8678 & 1.8   & 204 \\
    AlpScoB(13)I & 32.2  & 195.6 & 33.1  & 216.3 & 2447  & -1    & 3499  & 123.7631 & 1.8   & 321 \\
    AlpScoB(29)I & 32.2  & 286.5 & 65.7  & 343.9 & -2274.3 & 1     & 3452  & 62.04907 & 2.3   & 173 \\
    AlpVir(116)I & 17.3  & 243.3 & 34.7  & 188.8 & 5001  & -1    & 161196 & 55.39598 & 1.7   & 22 \\
    AlpVir(124)E & 17.3  & 123.3 & 29.5  & 218.4 & -2033.2 & 1     & 153398 & 203.5833 & 2.2   & 138 \\
    AlpVir(134)I & 17.3  & 284.9 & 50.4  & 215.4 & -637.2 & 1     & 158112 & 311.6297 & 2.2   & 346 \\
    AlpVir(173)E & 17.3  & 93    & 80.4  & 329.3 & -666.8 & -1    & 124007 & 200.0026 & 1.7   & 265 \\
    AlpVir(173)I & 17.3  & 31.2  & 20    & 71.1  & 2787.3 & -1    & 124993 & 200.0026 & 1.7   & 358 \\
    AlpVir(210)I & 17.3  & 311.5 & 25.1  & 46    & -4015.8 & 1     & 138215 & 43.22061 & 2.5   & 259 \\
    AlpVir(211)I & 17.3  & 267.2 & 81.2  & 103.3 & -2632.7 & 1     & 132499 & 469.4783 & 1.8   & 231 \\
    AlpVir(232)E & 17.3  & 89.3  & 87.8  & 316.7 & -2068.6 & 1     & 125750 & 455.5824 & 2.3   & 284 \\
    AlpVir(30)I & 17.3  & 230.4 & 26    & 6.9   & -2187.3 & 1     & 532397 & 177.7105 & 2.5   & 262 \\
    AlpVir(34)E & 17.3  & 332.9 & 19.2  & 259.5 & 2584  & 1     & 489542 & 247.7773 & 1.7   & 207 \\
    AlpVir(34)I & 17.3  & 232.8 & 27.2  & 24    & 579.6 & 1     & 510631 & 145.8352 & 2.2   & 225 \\
    AlpVir(8)E & 17.3  & 91.3  & 85.9  & 197.2 & 3679.9 & -1    & 526802 & 2500.032 & 1.7   & 20 \\
    AlpVir(8)I & 17.3  & 141.1 & 21.8  & 162.3 & 4488.6 & -1    & 500231 & 2499.995 & 1.8   & 145 \\
    BetCen(102)I & 66.7  & 249.1 & 81.3  & 0.9   & 4984.3 & -1    & 370169 & 329.3598 & 7.3   & 55 \\
    BetCen(104)E & 66.7  & 105.5 & 83.5  & 299.7 & -3402.1 & 1     & 342974 & 288.9692 & 8.6   & 155 \\
    BetCen(104)I & 66.7  & 209.3 & 69.4  & 321   & -1126.9 & -1    & 357501 & 312.9495 & 6.7   & 304 \\
    BetCen(105)E & 66.7  & 100.5 & 85.5  & 47    & -3710.5 & -1    & 287316 & 145.6972 & 6.3   & 353 \\
    BetCen(105)I & 66.7  & 212.9 & 70.1  & 83.1  & 5023  & 1     & 313435 & 151.4026 & 5.2   & 61 \\
    BetCen(64)E & 66.7  & 101.7 & 85    & 282.5 & 1854.5 & 1     & 619944 & 469.1418 & 5.6   & 286 \\
    BetCen(75)I & 66.7  & 270.5 & 89.8  & 232.9 & -934  & 1     & 594336 & 465.1562 & 7.6   & 273 \\
    BetCen(77)E & 66.7  & 47.5  & 73.8  & 12.6  & 4961  & 1     & 593240 & 336.3192 & 7.4   & 160 \\
    BetCen(77)I & 66.7  & 270.4 & 89.8  & 257.3 & -1823.4 & 1     & 587228 & 252.7589 & 7.7   & 165 \\
    BetCen(78)E & 66.7  & 45.9  & 73.3  & 206.5 & -5087.4 & -1    & 564075 & 375.1392 & 7     & 293 \\
    BetCen(81)I & 66.7  & 275.6 & 87.6  & 147.8 & 4636  & -1    & 550954 & 281.519 & 7.1   & 19 \\
    BetCen(85)I & 66.7  & 277.5 & 86.8  & 41.8  & -4526.9 & -1    & 533158 & 354.402 & 7.6   & 210 \\
    BetCen(89)I & 66.7  & 278   & 86.6  & 322.8 & -572.3 & 1     & 498234 & 330.1721 & 7.7   & 277 \\
    BetCen(92)E & 66.7  & 55    & 76.2  & 75.5  & 4143.3 & -1    & 452618 & 230.9221 & 6.2   & 319 \\
    BetCen(96)I & 66.7  & 271.9 & 89.2  & 126.3 & -4796.6 & -1    & 444264 & 229.5003 & 6.3   & 239 \\
    BetCMa(211)I & 14.2  & 223.4 & 19.2  & 67.1  & -3637.3 & 1     & 43518 & 50.86847 & 1.8   & 1 \\
    BetCru(253)I & 65.2  & 255.6 & 83.4  & 187.3 & -4002.7 & -1    & 106166 & 202.3216 & 6.5   & 157 \\
    BetCru(262)I & 65.2  & 257.3 & 84.2  & 18    & 5014.2 & -1    & 109532 & 185.7413 & 6.4   & 345 \\
    BetCru(98)I & 65.2  & 192.7 & 65.7  & 319   & 3954.6 & -1    & 274051 & 232.4198 & 7.1   & 248 \\
    BetPer(116)E & -47.4 & 143   & 53.7  & 216.7 & 1129.6 & 1     & 2135  & 43.00056 & 2.7   & 1 \\
    BetPer(116)I & -47.4 & 165.1 & 48.4  & 208.1 & -79.3 & 1     & 2130  & 43.00056 & 2.8   & 22 \\
    BetPer(42)I & -47.4 & 229.7 & 59.2  & 122.9 & -4479.8 & -1    & 20003 & 43.61236 & 3     & 244 \\
    ChiCen(39)I & 47.6  & 175.5 & 47.6  & 101.6 & 4977.8 & 1     & 13121 & 59.66402 & 3.8   & 345 \\
    DelCen(183)E & 55.6  & 149.3 & 59.5  & 177.7 & 3370.9 & 1     & 11397 & 217.076 & 4.6   & 5 \\
    DelCen(185)E & 55.6  & 150   & 59.3  & 44.2  & -1532 & -1    & 10924 & 200.8173 & 4.5   & 336 \\
    DelCen(191)E & 55.6  & 138.4 & 62.9  & 335.4 & -1934.9 & 1     & 10932 & 211.6986 & 4.6   & 31 \\
    DelCen(194)I & 55.6  & 294.7 & 74    & 270.1 & -5045.4 & 1     & 11891 & 22.14481 & 4.1   & 54 \\
    DelCen(64)E & 55.6  & 110.6 & 76.4  & 194.3 & 4773.2 & 1     & 52909 & 90.11278 & 5     & 327 \\
    DelCen(64)I & 55.6  & 124.8 & 68.6  & 210.7 & 3089.3 & 1     & 53150 & 90.11546 & 4.9   & 229 \\
    DelCen(66)I & 55.6  & 135.4 & 64    & 124   & 3831.5 & -1    & 58632 & 45.05706 & 4.7   & 311 \\
    DelCen(98)I & 55.6  & 211.3 & 59.6  & 54.8  & 402.3 & 1     & 34467 & 90.11549 & 5.8   & 100 \\ \\

       \end{tabular}%
       
\end{table}%

\begin{table}[htbp]
       \fontsize{6pt}{7pt}\selectfont
       \addtolength{\tabcolsep}{0pt}
       \centering
       \begin{tabular}{l|cccccccccc}
          Occultation & $B \, \textrm{[deg]}$ & $\phi \, \textrm{[deg]}$ & $B_{\mathrm{eff}} \, \textrm{[deg]}$ & $\textrm{Lon. \, [deg]}$ & $Z$ & $\textrm{sign}(\dot Z)$&$b$ & $I_0$ & $\tau_{\mathrm{max}}$ & $\Phi_{\mathrm{theory}} \textrm{[deg]}$ \\ \hline
 
    DelPer(36)E & -54   & 66.5  & 73.9  & 296   & 2858.5 & -1    & 13677 & 376.012 & 3.9   & 357 \\
    DelPer(37)I & -54   & 264.5 & 86    & 145.7 & 4558.5 & -1    & 13632 & 37.60231 & 5     & 29 \\
    DelPer(39)I & -54   & 264.6 & 86.1  & 45    & -4744 & -1    & 12741 & 38.59457 & 4.2   & 203 \\
    DelPer(41)I & -54   & 238.6 & 69.3  & 166.7 & -496  & -1    & 12185 & 63.56438 & 3.6   & 14 \\
    DelPer(60)I & -54   & 277.2 & 84.8  & 192.3 & 4814.5 & -1    & 11711 & 20.13449 & 3.6   & 16 \\
    DelSco(236)I & 28.7  & 263.2 & 77.9  & 10.8  & -5022.7 & -1    & 22981 & 647.6574 & 2.8   & 171 \\
    EpsCas(104)E & -70   & 122.3 & 79    & 123.6 & -4468.4 & 1     & 4557  & 37.14048 & 3.6   & 300 \\
    EpsCas(104)I & -70   & 187.8 & 70.1  & 152.8 & -2034.7 & -1    & 4586  & 37.14048 & 3.8   & 87 \\
    EpsCen(65)I & 59.6  & 226.6 & 68.1  & 343.1 & -1573.5 & -1    & 129715 & 127.6824 & 6.6   & 220 \\
    EpsCMa(173)I & 26    & 267   & 84    & 187.1 & 4165.8 & 1     & 40005 & 13.82051 & 2.8   & 341 \\
    EpsCMa(174)I & 26    & 267.1 & 84    & 47.5  & -3225.3 & -1    & 44805 & 15.57652 & 2.8   & 344 \\
    EpsCMa(276)E & 26    & 50.6  & 37.5  & 217.7 & 4192  & -1    & 69532 & 117.8264 & 2.6   & 109 \\
    EpsLup(36)E & 51    & 44.9  & 60.2  & 135.3 & 2840.1 & -1    & 32671 & 71.62883 & 5.7   & 100 \\
    EpsLup(37)E & 51    & 359.8 & 51    & 356.7 & -1132.5 & 1     & 31958 & 72.21977 & 3.7   & 135 \\
    EtaLup(34)E & 44.5  & 357.1 & 44.5  & 90    & -3167.9 & -1    & 46980 & 82.68831 & 4.2   & 353 \\
    EtaLup(34)I & 44.5  & 296.2 & 65.7  & 218.8 & 2575.6 & -1    & 47613 & 83.364 & 3.4   & 130 \\
    GamAra(37)I & 61    & 248.7 & 78.6  & 62.1  & 4684.1 & 1     & 26751 & 58.10375 & 4.2   & 133 \\
    GamCas(100)E & -66.3 & 72.5  & 82.5  & 93.5  & -3408.3 & 1     & 54160 & 93.44911 & 6.1   & 259 \\
    GamCol(173)E & -39.9 & 73.9  & 71.7  & 317.7 & 5020.4 & -1    & 1097  & 57.81785 & 1.9   & 226 \\
    GamCol(205)I & -42.6 & 260.4 & 79.7  & 269.1 & -2977.3 & -1    & 843   & 44.80865 & 2     & 354 \\
    GamGru(40)E & 35.1  & 193.1 & 35.8  & 187.8 & 4856.5 & -1    & 7292  & 122.9506 & 3.2   & 32 \\
    GamGru(41)E & 35.1  & 204.5 & 37.7  & 98.5  & -1794.7 & -1    & 7579  & 158.2192 & 2.6   & 122 \\
    GamGru(41)I & 35.1  & 282.7 & 72.6  & 98.2  & 4282.8 & -1    & 8154  & 166.3223 & 2.2   & 225 \\
    GamLup(30)E & 47.4  & 114.6 & 69    & 173.1 & -3263.2 & 1     & 77667 & 59.06165 & 4.6   & 303 \\
    GamLup(32)E & 47.4  & 33.8  & 52.6  & 158.4 & 5097.4 & 1     & 72586 & 72.70017 & 5.4   & 306 \\
    GamPeg(172)E & -20.3 & 74.1  & 53.5  & 209.7 & 5059.2 & -1    & 12683 & 12.42252 & 1.8   & 295 \\
    GamPeg(172)I & -20.3 & 36.7  & 24.7  & 283.4 & 4683.3 & 1     & 11491 & 12.42252 & 1.7   & 328 \\
    GamPeg(211)E & -20.3 & 122.5 & 34.6  & 207.2 & 518.8 & -1    & 13192 & 16.91014 & 1.8   & 200 \\
    GamPeg(32)I & -20.3 & 138.6 & 26.2  & 226.6 & 2857.3 & 1     & 74756 & 265.8262 & 1.8   & 343 \\
    GamPeg(36)E & -20.3 & 66.8  & 43.2  & 343.5 & 3206.1 & 1     & 70975 & 225.4201 & 1.8   & 60 \\
    KapCen(35)E & 48.5  & 85.5  & 86.1  & 302.8 & -5074.8 & 1     & 46015 & 57.75644 & 5.8   & 281 \\
    KapCen(36)I & 48.5  & 241.2 & 67    & 228.6 & -2480.8 & -1    & 44129 & 119.2304 & 5     & 330 \\
    KapCen(42)I & 48.5  & 168.5 & 49.1  & 237.3 & 228.4 & 1     & 41129 & 119.2288 & 4.1   & 88 \\
    KapCMa(168)E & 29.3  & 127.3 & 42.8  & 286.8 & -1613.5 & 1     & 6193  & 43.59726 & 1.7   & 229 \\
    KapCMa(168)I & 29.3  & 175.9 & 29.4  & 268.8 & -3886.5 & 1     & 6062  & 43.59726 & 1.7   & 270 \\
    KapSco(247)I & 43.4  & 273.4 & 86.4  & 326.7 & 4703.2 & -1    & 21356 & 409.719 & 3     & 21 \\
    LamSco(248)I & 41.7  & 283.9 & 75    & 261.3 & 585.4 & -1    & 58157 & 433.8854 & 3.7   & 343 \\
    LamSco(29)E & 41.7  & 148.3 & 46.3  & 222.1 & -4753.7 & -1    & 283952 & 159.058 & 4.6   & 35 \\
    LamSco(44)I & 41.7  & 235   & 57.2  & 62.7  & -841.5 & -1    & 252102 & 136.4883 & 6.1   & 254 \\
    MuCen(113)I & 48.7  & 239.1 & 65.8  & 24.1  & -3432.9 & -1    & 9633  & 23.36014 & 3.7   & 81 \\
    MuSco(43)E & 43.4  & 27    & 46.7  & 303.3 & 4005.9 & -1    & 86896 & 441.6879 & 4     & 130 \\
    PsiCen(38)I & 44.3  & 249.9 & 70.6  & 54.2  & 3180.3 & 1     & 1110  & 58.71554 & 2.2   & 137 \\
    SigSgr(11)I & 29.1  & 239.5 & 47.6  & 111   & -2828 & -1    & 118989 & 995.783 & 3.4   & 85 \\
    SigSgr(114)I & 29.1  & 330.2 & 32.6  & 262.3 & -492.7 & 1     & 33500 & 40.79908 & 2.7   & 160 \\
    SigSgr(244)E & 29.1  & 234.2 & 43.5  & 103.7 & 204.5 & 1     & 19555 & 40.79669 & 1.9   & 262 \\
    SigSgr(244)I & 29.1  & 269.3 & 88.8  & 86.6  & -1341.1 & 1     & 16595 & 40.79913 & 2.1   & 313 \\
    TheAra(40)E & 53.9  & 28.6  & 57.3  & 197.2 & -2036.3 & -1    & 12489 & 46.22136 & 4.1   & 270 \\
    TheAra(40)I & 53.9  & 356.2 & 53.9  & 317   & 4476.6 & -1    & 12490 & 46.22205 & 3.7   & 66 \\
         TheAra(41)E & 53.9  & 82.4  & 84.5  & 205.4 & 1720.2 & -1    & 11762 & 46.22204 & 5.5   & 194 \\
             TheCar(190)I & -43.3 & 252.4 & 72.2  & 219.3 & -4463.7 & 1     & 25334 & 159.9588 & 5.7   & 277 \\
    ZetCen(112)I & 53.6  & 239.7 & 69.6  & 112.5 & 4393.8 & -1    & 37307 & 34.73377 & 6.4   & 346 \\
    ZetCen(246)E & 53.6  & 79.3  & 82.2  & 254.4 & -4100.4 & 1     & 22256 & 232.1636 & 5.1   & 144 \\
    ZetCen(60)I & 53.6  & 228.1 & 63.8  & 258.7 & -4185.7 & -1    & 107181 & 113.7307 & 6.6   & 235 \\
    ZetCen(62)E & 53.6  & 70.1  & 75.9  & 8.9   & -4066 & 1     & 106091 & 96.60519 & 5.5   & 47 \\
    ZetPup(171)E & 38.6  & 117.3 & 60.1  & 154.4 & -2266.7 & -1    & 46835 & 141.573 & 4.5   & 264 \\
    ZetPup(171)I & 38.6  & 202.1 & 40.8  & 120.2 & 2022.7 & -1    & 49506 & 122.4082 & 4.3   & 351 \\
       \end{tabular}%
       \caption{Complete UVIS dataset for occultations with $\tau_{\mathrm{max}}  > 1.5$ in the Mimas 5:3 bending wave region.}
\end{table}%

\bibliography{arxiv}

\end{document}